\documentclass[journal=jacsat,manuscript=article]{achemso}

\usepackage[version=3]{mhchem} 
\usepackage{chemformula}

\author{Fabrizio Camerin}
\affiliation{Division of Physical Chemistry, Department of Chemistry, Lund University, P. O. Box 124, SE-22100 Lund, Sweden} 
\email{fabrizio.camerin@chem.lu.se}

\author{Megha Emerse}
\affiliation{Department of Chemistry, University of Florence, via della Lastruccia 3, Sesto Fiorentino, I-50019 Firenze, Italy}

\author{Gianpaolo Gallo}
\affiliation{Laboratory for Soft Materials and Interfaces, Department of Materials, ETH Zürich, Vladimir-Prelog-Weg 5, 8093 Zürich, Switzerland}

\author{Najet Mahmoudi}
\affiliation{ISIS Neutron and Muon Source, Rutherford Appleton Laboratory, Oxfordshire, OX11 0QX, UK}

\author{Gregory Smith}
\affiliation{ISIS Neutron and Muon Source, Rutherford Appleton Laboratory, Oxfordshire, OX11 0QX, UK}

\author{Emanuela Zaccarelli}
\affiliation{CNR Institute of Complex Systems, Uos Sapienza, Piazzale Aldo Moro 2, 00185, Roma, Italy}
\alsoaffiliation{Department of Physics, Sapienza University of Rome, Piazzale Aldo Moro 2, 00185 Rome, Italy}

\author{Lucio Isa}
\affiliation{Laboratory for Soft Materials and Interfaces, Department of Materials, ETH Zürich, Vladimir-Prelog-Weg 5, 8093 Zürich, Switzerland}

\author{Marco Laurati}
\affiliation{Department of Chemistry, University of Florence, via della Lastruccia 3, Sesto Fiorentino, I-50019 Firenze, Italy}
\alsoaffiliation{Consorzio interuniversitario per lo sviluppo dei Sistemi a Grande Interfase (CSGI), via della Lastruccia 3, 50019 Sesto Fiorentino (FI), Italy}
\email{marco.laurati@unifi.it}

\author{Jacopo Vialetto}
\affiliation{Department of Chemistry, University of Florence, via della Lastruccia 3, Sesto Fiorentino, I-50019 Firenze, Italy}
\alsoaffiliation{Consorzio interuniversitario per lo sviluppo dei Sistemi a Grande Interfase (CSGI), via della Lastruccia 3, 50019 Sesto Fiorentino (FI), Italy}
\email{jacopo.vialetto@unifi.it}

\title[An \textsf{achemso} demo]{Resolving Light-Induced Structural Rearrangements in Responsive Microgels} 

\abbreviations{IR,NMR,UV}

\begin{document}

%
%

\begin{abstract}
Optically-responsive microgels offer a versatile platform for designing adaptive soft materials with coupled light and thermal responsiveness. Control over the crosslinking degree is particularly appealing as it can regulate not only particle size but also stiffness, thereby enabling remote tuning of key material functionalities. However, the internal structural changes that couple molecular photoresponsive mechanisms to mesoscopic properties remain poorly resolved. Here, we investigate different light-responsive microgels containing covalently incorporated coumarin moieties, which impart optical sensitivity through UV-induced cycloaddition, by combining dynamic light scattering, small-angle neutron scattering, and molecular dynamics simulations. We show that light irradiation alters not only particle size but also the internal polymer density distribution and subsequent thermal response. Before irradiation, the microgels exhibit a star-like architecture with a dense core and extended polymeric arms. After irradiation, the network evolves toward a markedly more compact structure. This transformation cannot be rationalized simply as an equivalent to an increase in crosslinking density during synthesis, as observed in the thermal response, revealing light as a powerful tool to regulate microgel architecture and multifunctional responsiveness.
\end{abstract}


\section{Introduction}

Stimuli-responsive colloidal systems attract increasing interest due to their ability to dynamically alter their properties in response to external inputs such as temperature or pH variations, light or magnetic fields. \cite{Cao2016,Grzybowski2017} By coupling microscopic molecular switches to macroscopic material properties, they enable the study of phase transitions, \cite{Vialetto2019,Vialetto2021} self-assembly, \cite{Klajn2007} and non-equilibrium phenomena, \cite{Zhao2025} and allow designing smart materials with adaptive mechanical, optical, or rheological behaviors, which are of interest in areas ranging from photonics to biomedicine. \cite{Grzybowski2017}
Particularly appealing are particles able to respond independently to multiple stimuli, therefore allowing for a superior control over the final responsiveness and functionalities. Soft colloids, e.g. microgels, fall into this category. When made of thermoresponsive polymers, these particles switch from a swollen state at low temperature, when the polymer is in a good solvent condition, to a collapsed state above the volume phase transition temperature (VPTT). This temperature-dependent response is driven by changes of polymer-solvent interactions. \cite{Plamper2017} Thermoresponsive particles, such as those based on poly(N-isopropylacrylamide) (PNIPAM), can be precisely synthesized to control their size, shape and softness, and are endowed with multiple responsiveness either by complexation with molecules or nanoparticles, or by copolymerization with suitable monomers. \cite{Karg2019}

Among the various stimuli exploited to design adaptive materials, light has several advantages such as fast, contactless and precise spatio-temporal control, allowing rapid switching with superior spatial resolution. \cite{Mal2003,Jampani2024}
Light-responsiveness in microgels can be obtained by incorporating responsive nanoparticles, \cite{Gorelikov2004} or by introducing photo-reactive moieties, which are either non covalently \cite{Bradley2006,Jelken2022,Sharma2023} or covalently attached \cite{Garcia2007,He2009,Klinger2011,Phua2016,Lu2019,Lu2020,Hu2022,Dong2023,Agnihotri2024} within the polymeric network.
In particular, molecular switches can be used to tune the crosslinking density within the microgels \textit{post-synthesis}. In this way, they can degrade the particle on-demand when photo-cleavable crosslinkers are introduced \cite{Klinger2011}, or they can modulate the swelling-deswelling ability of the microgel \cite{Zhang2014,Lu2019,Liang2022} and the overall mechanical properties. \cite{Lu2020} This is of great interest for the design of novel materials and devices since microgels' softness is the main parameter that imparts them with superior functionalities with respect to mechanically rigid colloids. \cite{Brijitta2019}
To date, most studies have primarily focused on the synthesis and macroscopic characterization of light-responsive microgels incorporating a wide range of photo-switches. \cite{Shu2022} 
These works have been essential in demonstrating the feasibility of designing microgel networks whose properties can be modulated by light. However, considerably less attention has been devoted to resolving how light irradiation affects the internal architecture of these soft colloidal networks. In particular, the conformational rearrangements occurring within the microgel interior, and how these depend on the polymeric composition and photo-switch distribution remain largely unexplored. A more detailed understanding of these internal structural changes is crucial for establishing a direct connection between molecular-level photoresponsive mechanisms and the resulting mesoscopic properties of the microgel. 
This information will then be of practical interest in light-switchable materials, where both global or local properties, such as stiffness and interpenetration, can be modulated by applying spatially controlled optical stimuli.

Here, we investigate the internal polymer and crosslinker distribution in responsive microgels by combining dynamic light scattering (DLS), small-angle neutron scattering (SANS) and numerical simulations, unveiling detailed insights into their structural properties and overall size upon combined thermal and light actuation.
The microgels are synthesized with triethylene glycol methyl ether methacrylate (MeO$_3$MA) as the main monomer, similarly to the works of Lu et al.~\cite{Lu2019,Lu2020}, and compared with microgels containing the more common NIPAM monomer. As crosslinker we used ethylene glycol dimethacrylate (EGDMA) which, when mixed with NIPAM, results in the formation of so-called star-like thermoresponsive microgels with a densely crosslinked core and loose polymer arms~\cite{Ballin2025,Vialetto2026}. 
To impart light responsiveness, we covalently link the polymeric network with a modified coumarin comonomer. Coumarins are a class of photo-reactive molecules able to undergo [2+2] cycloaddition reactions upon irradiation in the UV region. \cite{Mal2003} The formation, or cleavage, of coumarin dimers has been used to remotely control the size of microgels~\cite{Lu2019} as well as the mechanical properties, healing and reshaping abilities or fluorescence response of hydrogels-containing responsive microgels. \cite{Lu2020,Dong2023}
Notably, we demonstrate that irradiation not only 
affects the microgel size, but also modulates the internal polymer density and, consequently, their thermal responsivity. Supported by molecular dynamics simulations, we show that light exposure induces a transition from an initially star-like polymer density profile to a significantly more compact architecture. Importantly, this structural transformation cannot be described as comparable to a simple increase in crosslinker content during synthesis, as evidenced by the distinctly altered thermoresponsive behavior observed after irradiation. Overall, we elucidate how the polymer network responds upon the application of multiple stimuli, linking architecture changes with microgel properties of interest for multiple applications. Our findings point at light as a versatile tool for controlling size, polymer density profile and combined thermal and optical responsiveness.

\section{Experimental}

\subsection{Reagents}
4-methylumbelliferone ($\geq$ 98 \%), methacryloyl chloride ($\geq$ 97.0 \%), ethylene glycol dimethacrylate (EGDMA, 98\%), ammonium persulfate (APS, 98\%), triethylene glycol methyl ether methacrylate (MeO$_3$MA, average $\text{M}_{\text{n}}$ = 200, 93\%), oligo (ethylene glycol) methyl ether methacrylate (OEGMA, average $\text{M}_{\text{n}}$ = 950 g/mol), methacrylic acid (MAA, 99\%), N-isopropylacrylamide (NIPAM, 99\%) and sodium dodecyl sulfate (SDS, 99\%) were purchased from Merck. 
2-bromoethanol ($>$ 95.0 \%) was purchased from Tokyo Chemical Industry, triethylamine (TEA, $>$ 99 \%) from VWR, extra-dry N,N-dimethylformamide (DMF, 99.8 \%) from Acros Organics.
MeO$_3$MA and MAA were purified before usage by filtration on alumina (Aluminum oxide, activated, neutral, Brockmann Grade II, Thermo Scientific Chemicals). NIPAM was purified by recrystallization in 40/60 v/v toluene/hexane.
Water used for all experiments is bi-distilled Milli-Q.

\subsection{CMA synthesis}
The procedure was adapted from Refs.~\cite{Kabb2018,Lu2019}.

\noindent \textit{Step 1: synthesis of 7-(2-hydroxyethoxy)-4-methylcoumarin.} In a two-neck 100mL round-bottom flask equipped with a reflux condenser, 4-methylumbelliferone (4.00 g, 22.7 mmol) and K$_2$CO$_3$ (6.23 g 45.4 mmol) were suspended in dry DMF (40 mL). The suspension was stirred and heated up to 90 °C, under constant N$_2$ flow. After this temperature was reached, 2-bromoethanol (2.42 mL, 34.0 mmol) was added dropwise to the reaction mixture. The reaction was then refluxed for 18 h at 90°C, then it was allowed to cool down to room temperature. The content of the reaction flask was then poured into ice-cold water (800 mL) and stirred inside an ice-bath for another 6 h. The mixture was filtered by vacuum filtration and the resulting white powder was dried under vacuum for three days. \\ 
\textbf{$^1$H NMR} (300 MHz, d6-DMSO) $\delta$ 7.72 – 7.62 (m, 1H), 7.03 – 6.92 (m, 2H), 6.20 (q, J = 1.2 Hz, 1H), 4.99 – 4.88 (m, 1H), 4.09 (dd, J = 5.4, 4.4 Hz, 2H), 3.74 (q, J = 4.6 Hz, 2H), 2.39 (d, J = 1.3 Hz, 3H).

\noindent \textit{Step 2: synthesis of 2-((4-methyl-2-oxo-2H-chromen-7-yl)oxy)ethyl methacrylate
(CMA).} In a 250 mL round bottom flask placed in an ice-bath, 7-(2-hydroxyethoxy)-4-methylcoumarin (4.50 g, 20.4 mmol) was suspended in chloroform (72 mL). Triethylamine was added through a septum while stirring (4.50 g, 44.2 mmol). After 10 minutes, dropwise addition of methacryloyl chloride (4.50 g, 43.0 mmol) into the reaction mixture was performed at 0 °C. The solution was left to warm up to room temperature and then stirred overnight. The reaction mixture was diluted with dichloromethane (DCM), washed with brine three times and then dried over anhydrous MgSO$_4$. The latter was removed by gravity filtration and the resulting liquid concentrated under reduced pressure. The crude product was purified by double recrystallization from ethanol and it was dried overnight under vacuum. A yellow powder of pure 2-((4-methyl-2-oxo-2H-chromen-7-yl)oxy)ethyl methacrylate (CMA) was obtained. \\
\textbf{$^1$H NMR} (300 MHz, CDCl$_3$) $\delta$ 7.50 (d, J = 8.8 Hz, 1H), 6.93 – 6.78 (m, 2H), 6.14 (d, J = 1.5 Hz, 2H), 5.60 (p, J = 1.6 Hz, 1H), 4.53 (dd, J = 5.6, 4.0 Hz, 2H), 4.32 – 4.22 (m, 2H), 2.40 (d, J = 1.2 Hz, 3H), 1.95 (t, J = 1.2 Hz, 3H).

\subsection{Microgel synthesis}
The microgels used in this work were synthesized by radical precipitation polymerization. For the synthesis of microgels with MeO$_3$MA as the main monomer, CMA at the required mol \% (calculated with respect to all monomers present in the synthesis) was dissolved in a mixture of MeO$_3$MA (0.75 g), EGDMA (1 mol \%) and MAA (20 mol \%) by mixing at 65°C for one hour. In a round-bottom flask, OEGMA (4.9 mol \%) and SDS (0.2 mg/mL) were dissolved in 76 mL of Milli-Q water, purged with nitrogen for 1 h and placed at 80°C. To initiate the reaction, 1.4 mg of APS dissolved in 4 mL of Milli-Q water and purged with nitrogen for 1 h was added to the reaction flask immediately followed by the addition of the solution of CMA in the monomers mixture. Particular care in keeping the CMA solution above 60°C was required in order to avoid CMA precipitation. 
The reaction was carried out for 6 h and stopped by opening the flask to air and placing it in an ice bath. 

For the synthesis of microgels with NIPAM as main monomer, two different procedures were used.
For the synthesis of microgel $\mu G_{nip1}$, CMA (5 mol \%) was dissolved in a mixture of EGDMA (1 mol \%),  MAA (20 mol \%) and OEGMA (4.9 mol \%) by mixing at 65°C for one hour. In a round-bottom flask, NIPAM (0.5 g) and SDS (0.2 mg/mL) were dissolved in 96 mL of Milli-Q water, and the reaction was then carried out as for MeO$_3$MA microgels.
For the synthesis of microgel $\mu G_{nip2}$, NIPAM (0.5 g), EGDMA (1 mol \%), OEGMA (3.2 mol \%) and CMA (5 mol \%) were mixed in 136 mL of Milli-Q water in a round-bottom flask. After purging with nitrogen for 1 h at 80°C, 1.4 mg of APS dissolved in 4 mL of Milli-Q and previously purged with nitrogen was added to initiate the reaction, which was carried out for 6 h and stopped by opening the flask to air and placing it in an ice bath. 

After the synthesis, all polymeric dispersions were purified through dialysis using a membrane with a molecular weight cut-off of 12-14 kDa, with a constant exchange of Milli-Q water for 10 days minimum. 

\begin{table}
  \caption{Microgels synthesis, hydrodynamic diameter from DLS at 15°C, deswelling ratio}
  \label{tbl:synthesis}
  \begin{tabular}{llllll}
    \hline
    Microgel & main monomer & CMA \% & D$_h$ (15°C) [nm] & \(D_h(47^\circ C)/D_h(15^\circ C)\) & VPTT [°C] \\
    \hline
$\mu G_{2.5}$  & MeO$_3$MA & 2.5 & 135 $\pm$ 5 & 0.74 $\pm$ 0.04 & $>$ 50 \\    
$\mu G_{5}$    & MeO$_3$MA & 5   & 149 $\pm$ 1 & 0.60 $\pm$ 0.01 & 32  \\    
$\mu G_{10}$   & MeO$_3$MA & 10  & 194 $\pm$ 2 & 0.56 $\pm$ 0.01 & 26 \\    
$\mu G_{nip1}$  & NIPAM    & 5  & 211 $\pm$ 3 & 0.63 $\pm$ 0.01 & 36 \\    
$\mu G_{nip2}$  & NIPAM    & 5  & 282 $\pm$ 2 & 0.71 $\pm$ 0.01 & 33 \\    
    \hline
  \end{tabular}
\end{table}

\subsection{Experimental methods}

\noindent \textit{$^1$H NMR.} $^1$H NMR spectra were recorded at room temperature on a Bruker UltraShield 300 MHz machine and were referenced to residual solvent peaks of CDCl$_3$ or d6-DMSO. The values for the coupling constants were determined assuming first-order coupling and the multiplicities are abbreviated as s (singlet), d (doublet), t (triplet), q (quartet), p (quintet), hept (heptet), m (multiplet) and combinations thereof. \\

\noindent \textit{Dynamic light scattering.} DLS experiments were performed using a Zetasizer PRO Red Label (Malvern, UK). The microgels were redispersed in Milli-Q water at a concentration of 0.01 wt\% in order to analyze samples in the dilute regime.
The temperature was varied from 15 to 47 or 51°C with 1°C or 2°C steps and 15 min equilibration time. For each temperature we recorded three consecutive measurements of 20 runs each. \\

\noindent \textit{Small-angle neutron scattering.} SANS measurements were performed at ISIS Neutron and Muon Source (RAL, Didcot, UK) using SANS2D and Larmor beamlines.
At SANS2D we used a a polychromatic beam with a wavelength range of $ 1.75<\lambda<12.5 $ \AA~and a sample-to-detector distance of 5 and 12 m for the front and rear detector, respectively. The combination of these configurations gives a wave vector range $0.0015 \text{ \AA}^{-1} < Q < 0.50\text{ \AA}^{-1}$.
At Larmor, a simultaneous Q-range of $0.004 \text{ \AA}^{-1} < Q < 0.68\text{ \AA}^{-1}$ was achieved utilizing an incident wavelength range of 0.9–13 \AA~and employing an instrument setup of L1 = L2 = 4 m. All measurements were performed at T = 20.0 $\pm$ 0.1 °C unless otherwise stated, in quartz cells with a path length of 2 mm (Hellma GmbH \& Co., Mullheim, Germany). Microgel suspensions at 0.1 w/w \% were prepared using a mixture of 9:1 D$_2$O-H$_2$O as solvent for MeO$_3$MA microgels, pure D$_2$O for NIPAM microgels.
All scattering data were normalized (for the transmission, thickness and detector efficiencies) and absolutely scaled using a standard polymer, and background corrected subtracting a quartz cell filled with the solvent, using MANTID \cite{ARNOLD2014156}. All data were analyzed within the SASView package using standard and user-written functions. \\

\noindent \textit{Light irradiation.} Irradiation was performed with a custom-made setup using a Mounted LED @ 340 nm (M340L5, Thorlabs GmbH, Germany) 
and a UV-grade fused silica lens for collecting the light in a disk of approximately 7 cm in diameter at the sample plane. For irradiation at 340 nm, dilute microgel suspensions were sealed in quartz cells and placed horizontally on a hot plate set at 40 °C, at a distance of 10 cm from the LED source mounted above it. The LED was used at its maximum power of 69.2 mW. 
The LED was used at its maximum power of 55.7 mW.

\noindent \textit{SANS data analysis.} Fitting of the microgel's form factor was performed with SasView software by using the star-like core-shell model from: \cite{Ballin2025,Vialetto2026}
\begin{equation}\label{eq:core-fuzzy-shell}
P(Q)= A_1(\frac{3}{V_t})(V_c(\rho_c - \rho_s)A_c + V_t(\rho_s - \rho_{solv})A_p)^2 + A_2\frac{\sin(\mu\tan^{-1}(Q\xi))}{Q\xi (1+Q^{2}\xi^{2})^{\mu/2}}
\end{equation}
In eq. \ref{eq:core-fuzzy-shell}, the the first term describes a core of radius $R_c$, scattering length density $\rho_c$ and volume $V_c = \frac{4}{3}\pi R_c^2$, and a fuzzy shell of thickness $t$, fuzziness $\sigma$ and scattering length density $\rho_s$. The core plus shell radius is $R_s = R_c + t$, while the total radius is $R_{tot} = R_s + 2\sigma$. $\rho_{solv}$ is the scattering length density of the solvent. The core is described by the scattering term:
\begin{equation}
   A_c = \frac{[\sin(QR_c) - QR_c\cos(QR_c)]}{(QR_c)^3}
\end{equation}
while the whole particle by a scattering term characteristic of a fuzzy sphere:
\begin{equation}
    A_p = \frac{[\sin(QR_s) - QR_s\cos(QR_s)]}{(QR_s)^3} \cdot \exp{\left( - \frac{(\sigma Q)^2}{2} \right)}
\end{equation}
The second part in eq. \ref{eq:core-fuzzy-shell} models the blob scattering of the star arms through the blob size $\xi$ representative of the characteristic length scale at which the granular polymer structure becomes relevant. The exponent $\mu$ is defined as $\mu=1/\nu-1$, with $\nu$ the Flory exponent. The contributions of the two terms in eq. \ref{eq:core-fuzzy-shell} is weighted by the amplitudes $A_{1}$ and $A_{2}$.

\subsection{Numerical methods}

\noindent \textit{Model for light-responsive microgels.} In our simulations, we use coarse-grained bead-based microgels assembled \textit{in silico} according to the procedure described in Refs.~\cite{Ballin2025,papetti2026star}, which allows us to obtain star-like microgels, with a tiny central core surrounded by long, crosslinker-free arms. Each bead has
unitary diameter $\sigma$ and mass $m$. To best reproduce experimental data, the EGDMA crosslinker concentration is set equal to $\approx 0.5\%$. Monomers mimicking coumarin are randomly assigned among the microgel beads i) within a varying radial distance from the center of mass (up to $\approx 40, 50$ and $55\%$ of the microgel extension), ii) randomly across the whole structure and iii) in an external shell (from $\approx 70\%$ of the overall extension outwards), depending on the case studied. In all cases, a fraction of $10\%$ of microgel beads is chosen to be coumarin. While interactions do not vary (see below), coumarin beads can only establish bond among themselves, which is allowed if their distance is smaller than $1.1\sigma$. Contrary to laboratory microgels, we do not consider the presence of other kinds of monomers and the addition of charges to the model. The numerical microgels used in this work have a nominal number of monomers $N=42k$, assembled in a sphere of radius $Z=50\sigma$. To account for the effect of polydispersity, we also synthesized microgels with $N=29,38,46$ and $58k$ monomers keeping the monomer density fixed and equal to $0.08\sigma^{-3}$. 

\noindent \textit{Interaction potentials.} Within microgels, all beads interact according to the Kremer-Grest bead-spring model, where all monomers experience a Weeks-Chandler-Anderson ($V_{\rm WCA}$) potential~\cite{grest1986}
\begin{equation}
V_{\rm WCA}(r)=
\begin{cases}
4\epsilon\left[\left(\frac{\sigma}{r}\right)^{12}-\left(\frac{\sigma}{r}\right)^{6}\right] + \epsilon & \text{if $r \le 2^{\frac{1}{6}}\sigma$}\\
0 & \text{otherwise}
\end{cases}
\end{equation}
where $\epsilon$ sets the energy scale and $r$ is the distance between two monomers. In addition, connected beads also interact via the Finitely Extensible Nonlinear Elastic (FENE) potential~\cite{warner1972},
\begin{equation}
V_{\rm FENE}(r)=-\epsilon k_FR_0^2\ln\left[1-\left(\frac{r}{R_0\sigma}\right)^2\right]     \text{ if $r < R_0\sigma$,}
\end{equation}
with $k_F=15$ determining the stiffness of the bond and $R_0=1.5$ the maximum bond distance. 
To account for coumarin being hydrophobic, we consider an additional solvophobic potential~\cite{soddemann2001generic} among coumarin beads:
\begin{equation}
  \label{va_force}
  V_\alpha(r)=
  \begin{cases}
    -\epsilon \alpha \ \ \        &\text{if}\ r\leq 2^{1/6}\sigma,\\
    \frac{1}{2}\alpha \epsilon  \big[ \cos(\gamma(r/\sigma)^2+\beta)-1 \big]  \ \ \     &\text{if}\ 2^{1/6}\sigma<r<R_0\sigma,\\
    0 \ \ \ &\text{if}\ r>R_0\sigma.
  \end{cases} 
\end{equation}
Here $\gamma=\pi (2.25-2^{1/3})^{-1}$, $\beta = 2\pi-2.25\gamma$ and $\epsilon$ is the unit of energy. This potential effectively exerts an attraction between beads, thus reproducing an enhanced affinity. Here, we set $\alpha=0.8$.

\noindent \textit{Simulation details.} Simulations are carried out in the $NVT$ ensemble with the Nosè-Hoover thermostat~\cite{nose1984unified} setting the reduced temperature $T^*=k_BT/\epsilon=1$, with $k_B$ the Boltzmann constant and $T$ the temperature. We use \textsc{lammps} ~\cite{thompson2022lammps} for all simulations.
Each microgel is placed in the center of a cubic simulation box with side $250\sigma$ and periodic boundary conditions. After the assembly, we carry out an equilibration run for at least $1\times10^6 \delta t$, with $\delta t=0.002\tau$ and $\tau=\sqrt{m \sigma^2/\epsilon}$ the unit of time. Subsequently, for studying coumarin-crosslinked microgels, we let coumarin monomers in microgels to form bonds for at least $2\times10^7 \delta t$ after which we interrupt the formation of bonds and let the simulation to run for other $5\times10^6 \delta t$.

\noindent \textit{Analysis of the simulation trajectories.} We calculate the \textit{in silico} microgel form factor as
\begin{equation}
P(q)=\frac{1}{N}\sum_{i,j=1}^N \langle \exp{(-i\vec{q} \cdot \vec{r}_{ij})} \rangle,
\end{equation} 
where $\vec{r}_{ij}$ is the distance between monomers $i$ and $j$, while the angular brackets indicate an average over different configurations and over different orientations of the wavevector $\vec{q}$. The radial density profile of the microgel monomers is defined as 
\begin{equation}\label{eq:profile}
\rho(r)= \left\langle \frac{1}{N}\sum_{i=1}^{N} \delta (|\vec{r}_{i}-\vec{r}_{CM}|-r) \right\rangle,
\end{equation} 
\noindent
with $\vec{r}_{CM}$ the position of the center of mass of the particle. The sum runs either on all microgel particles or on coumarin beads only, depending on the density profiles that has to be obtained.

\section{Results and discussion}

\begin{figure}[t!]
\centering
\includegraphics[scale=1]{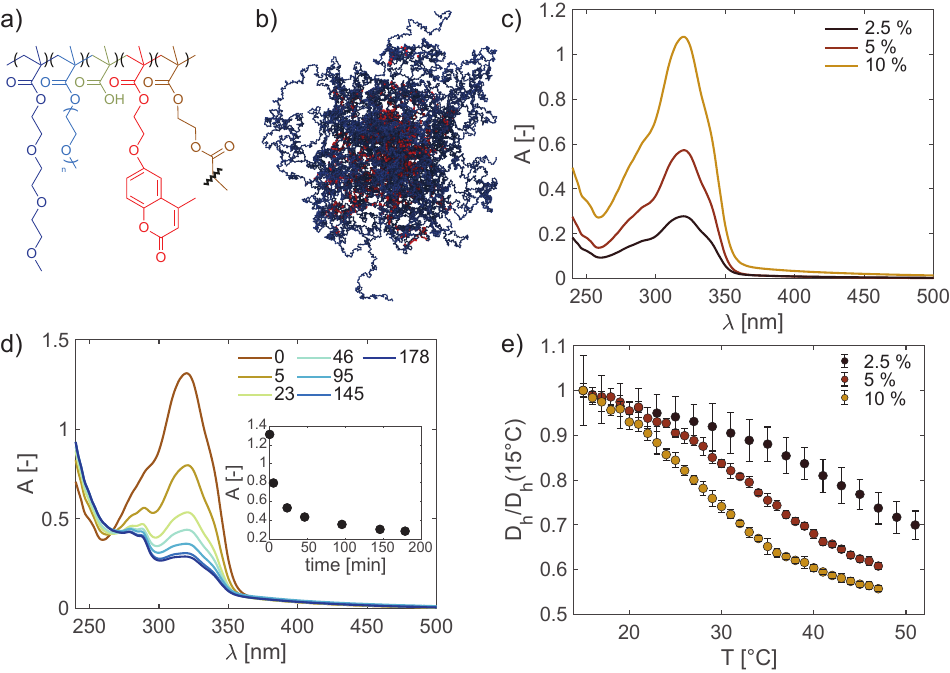}
\caption{\small Light-responsive microgels. a) Schematic of the polymer chain. b) Simulation snapshot of a microgel with coumarin units in red and non-coumarin beads in blue. c) UV-vis spectra of $\mu G_{2.5}$, $\mu G_{5}$ and $\mu G_{10}$ at a concentration of 0.18 mg/mL in MilliQ water. d) Spectral evolution of $\mu G_{5}$ upon irradiation at 340 nm for increasing time in minutes as indicated. Inset: absorbance value at 320 nm over time. e) Relative microgel hydrodynamic diameter $D_h$ in water divided by $D_h$ at 15 °C, as a function of temperature.}
\label{fig1}
\end{figure}

We first discuss experimental results on microgels with MeO$_3$MA as the main monomer, named $\mu G_{2.5}$, $\mu G_{5}$ and $\mu G_{10}$ according to the mol \% of the added coumarin comonomer (2-((4-methyl-2-oxo-2H-chromen-7-yl)oxy)ethyl methacrylate, CMA). 
The chemical structure of the resulting polymer network is reported in Figure~\ref{fig1}a. We note that multiple monomer species are employed, adapting the protocol described in Refs.~\cite{Lu2019,Lu2020}, in order to dissolve CMA in the hot liquid comonomers prior to initiating the synthesis (see additional details in the Experimental section). 
In numerical simulations, we model light-responsive microgels through a simplified representation compared to the experimental system (see Figure~\ref{fig1}b), by introducing a generic monomer, shown in blue, and a coumarin-like monomer, shown in red, whose spatial distribution within the network is explicitly controlled, and which can participate in bond formation.

UV-vis spectroscopy of the purified colloids obtained after dilution in water at equal microgel mass concentration (0.18 mg/L) are reported in Figure \ref{fig1}c. The peak at $\lambda \simeq$ 320 nm is indicative of coumarin moieties, with an increase in the absorbance value that indicates a higher incorporation of the coumarin derivative inside the polymer network depending on the CMA mol \% in the reaction mixture. 
CMA [2+2] cycloaddition is promoted by irradiating the samples at 40°C with a LED source with wavelength $\lambda = 340$ nm (see additional details in the Experimental section). Such a temperature is chosen, in accordance with the literature, to trigger the cycloaddition in polymer networks in a collapsed or partially collapsed state. \cite{Lu2019} UV-vis measurements are used to quantify the dimerization kinetic by following the spectral evolution as a function of irradiation time with the representative microgel $\mu G_{5}$ (Figure \ref{fig1}d). The intensity decrease of the absorbance band around 320 nm is quantified in the inset in Figure \ref{fig1}d. This confirms that photo-crosslinking is taking place and it is completed within less than 200 minutes of irradiation time.

Dynamic light scattering (DLS, Figure \ref{fig1}e) confirms the successful synthesis of monodisperse particles. The hydrodynamic diameter $D_h$ in the swollen state at 15°C increases with increasing CMA content (see Table \ref{tbl:synthesis} of the Experimental section), similarly to what was observed by Lu and coworkers. \cite{Lu2019} We attribute this result to a decreased stabilization of the growing particles during synthesis in the presence of hydrophobic CMA, giving rise to larger polymeric networks. \cite{Gan2002} All microgels maintain their thermoresponsive behavior, with a broad transition from swollen to collapsed state that is dependent on the \% of CMA. In particular, the VPTT, measured as the inflection point of a sigmoidal fit of $D_h$ as a function of temperature, decreases from above 50°C for $\mu G_{2.5}$, to 32°C and 26°C for $\mu G_{5}$ and $\mu G_{10}$, respectively. Interestingly, the deswelling ratio (\(D_h(47^\circ C)/D_h(15^\circ C)\)) is also dependent on the CMA content and decreases from 0.74 $\pm$ 0.04 for $\mu G_{2.5}$ to 0.56 $\pm$ 0.01 for $\mu G_{10}$ (Table \ref{tbl:synthesis}). The fact that both the VPTT and the deswelling ratio decrease with increasing CMA can be attributed to the hydrophobicity of the coumarin comonomer, which increases the overall hydrophobicity of the polymer network, shifting the VPT to lower temperatures, and causing formation of a more compact and dense colloid at high temperature. Conversely, microgels with lower CMA content maintain more water inside the network, causing a lower deswelling for comparable temperature (here 47°C).

To expand the library of available monomers for light-responsive microgels synthesis, we then prepared particles containing NIPAM as the main monomer by following two procedures. 
For $\mu G_{nip1}$, in analogy with the $\mu G$ series based on MeO$_3$MA, CMA was first dissolved in the hot liquid comonomer mixture, containing EGDMA, MAA, and OEGMA, as detailed in the Experimental section. This solution was then added to the reaction flask immediately before initiating the polymerization.
Instead, for $\mu G_{nip2}$ CMA was simply added in water together with the other monomers. The second protocol is generally avoided as CMA is poorly soluble in water, and small aggregates remain visible when the synthesis begins.
DLS analysis of the resulting particles reveals that both synthesis protocols yield monodisperse microgels, with a VPTT of 36°C for $\mu G_{nip1}$ and 33°C for $\mu G_{nip2}$, and a deswelling ratio of 0.63 $\pm$ 0.01 and 0.71 $\pm$ 0.01 for $\mu G_{nip1}$ and $\mu G_{nip2}$, respectively (Table \ref{tbl:synthesis} and Figure S1). A lower deswelling for $\mu G_{nip2}$ could be attributed to the lower amount of hydrophilic OEGMA added during synthesis, as well as to the different protocol used. In both cases the particles maintain their thermal responsiveness.

\begin{figure}[t!]
\centering
\includegraphics[scale=0.9]{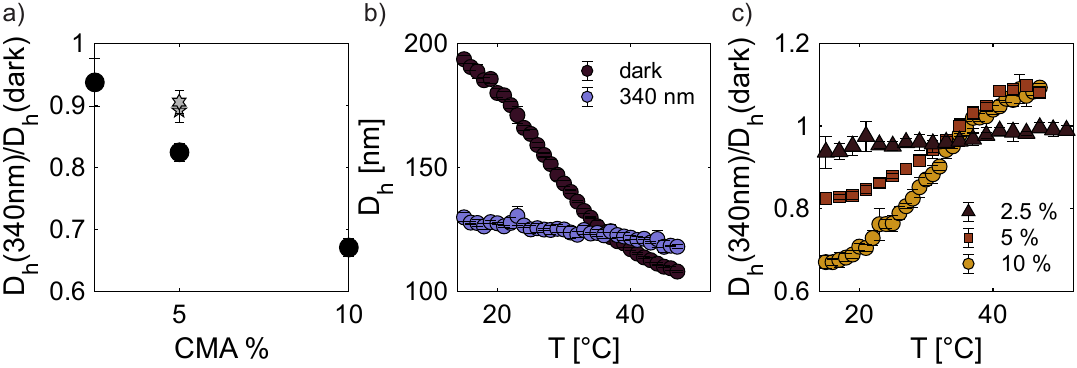}
\caption{\small Particle size upon irradiation at 340 nm. a) Relative microgel hydrodynamic diameter after irradiation $D_h (340nm)$ divided by $D_h$ in dark, as a function of CMA content. Dark circles: $\mu G$ microgels; pentagram: $\mu G_{nip1}$; hexagram: $\mu G_{nip2}$ b) $D_h$ as a function of temperature for $\mu G_{10}$ in dark and after irradiation at 340 nm. c) Relative size $D_h (340nm) / D_h (dark)$ as a function of temperature for $\mu G$ microgels containing different amount of CMA.}
\label{fig2}
\end{figure}

DLS is used to analyze the influence of dimerization upon extensive irradiation at $\lambda = 340$ nm on the network size and thermal responsivity. $D_h$ measured in the swollen state at 15°C after irradiation reveals that the microgels' size is light-sensitive. Binding of CMA units within the collapsed polymer network at 40°C causes a decreased reswelling of the irradiated particles with respect to the dark state due to the presence of additional crosslinking points. The extent of light-responsiveness is dependent on the CMA content, with a \(D_h(340nm)/D_h(dark)\) ratio (Figure \ref{fig2}a) that decreases from 0.93 $\pm$ 0.04 for $\mu G_{2.5}$, indicating limited light responsiveness with 2.5 mol \% CMA, to 0.82 $\pm$ 0.01 and 0.67 $\pm$ 0.01 with increasing CMA.
NIPAM microgels, both containing 5 mol \% CMA, show instead a \(D_h(340nm)/D_h(dark)\) ratio of 0.89 $\pm$ 0.02 and 0.91 $\pm$ 0.02 for $\mu G_{nip1}$ and $\mu G_{nip2}$, respectively (Figure \ref{fig2}a), demonstrating that the particles maintain light-responsiveness despite the different monomers forming the network. The slightly lower variation observed in $\mu G_{nip2}$ with respect to $\mu G_{nip1}$, and even more to $\mu G_{5}$, can be attributed to the lower incorporation of CMA during synthesis due to its limited solubility in the reaction medium.

We then investigated the behavior of the irradiated microgels as a function of temperature. After irradiation at 340 nm, $\mu G_{2.5}$ maintains a similar thermal responsiveness as the dark state, and the particle size at high temperature is comparable (Figure S2). A similar result is also obtained for $\mu G_{nip1}$ and $\mu G_{nip2}$ (Figures S3 and S4), which display a decreased light-responsiveness with respect to the microgel containing MeO$_3$MA and the same amount of CMA in the reaction flask.
Instead, a higher CMA content inside the network causes a decrease of the thermal responsiveness (Figure S5 for $\mu G_{5}$ and Figure \ref{fig2}b for $\mu G_{10}$) as expected due to the significant increase of effective crosslinking points, which hinders swelling at low temperature. 
Figure \ref{fig2}c reports the microgel size after irradiation with respect to the initial size \(D_h(340nm)/D_h(dark)\) for each temperature, capturing the limited light-sensitivity of $\mu G_{2.5}$ and the increased light-responsiveness of $\mu G_{5}$ and $\mu G_{10}$, which is also temperature-dependent.
Surprisingly, the particle size at high temperature is higher than the size of the non-irradiated microgels. This somewhat counterintuitive behavior can be rationalized considering that CMA binding differs from the addition of crosslinker during synthesis as the novel crosslinking points link already formed polymer chains in a partial or completely collapsed and entangled state at 40°C. These additional crosslinking points limit the ability of the polymer network to rearrange and either incorporate or expel solvent.
\begin{figure}[t!]
\centering
\includegraphics[scale=0.9]{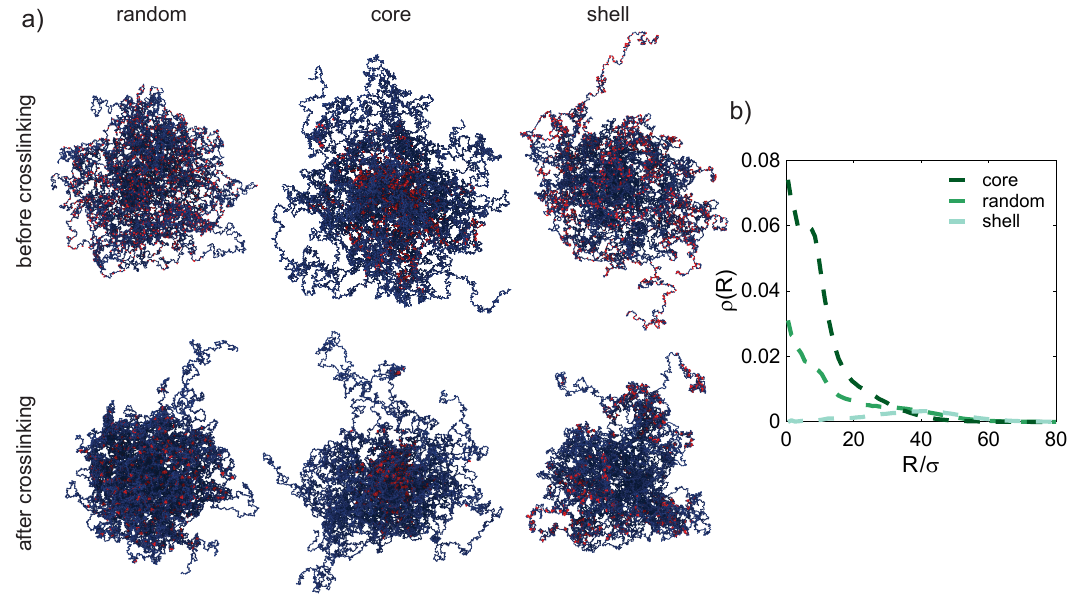}
\caption{\small Different coumarin distribution within the microgel network. a) Representative simulations snapshots showing three different distribution of coumarin (in all cases containing 10 \% coumarin): randomly in whole structure of the particle, mainly in the core, and only in an external shell. These distributions are named as random, core and shell, respectively, and are shown both before (upper panels) and after (lower panels) crosslinking. Coumarin monomers are slightly enlarged compared to their effective size for visual clarity. (b) Radial density profiles of coumarin monomers only, for the microgels before crosslinking shown in a). 
}
\label{fig3}
\end{figure}
\\
\\
\indent The structural changes observed in the microgel are expected to arise from the interplay between the organization of coumarin within the network and its responsiveness to light irradiation. To elucidate these effects, we adopt a minimal coarse-grained description of microgels, where each bead represents a monomer. Specifically, we build on a previously validated star-like microgel architecture, shown to reproduce experimentally synthesized microgels with high accuracy~\cite{Ballin2025}. Within this framework, different coumarin distributions can be explored in a direct and controlled manner by assigning bond-forming capability to selected monomers during a molecular dynamics simulation. Additionally, we account for the different degree of hydrophobicity of coumarin compared to other monomers by means of a solvophobic interaction potential. More details on modeling and simulations  can be found in Methods. 
In this way, we generate three different coumarin distributions representing three limiting cases of spatial organization within the microgel: i) a random distribution throughout the whole network, independently of the distinction between inner core and outer shell; ii) a distribution mainly concentrated in the core region; and iii) a distribution restricted to the periphery, giving rise to a responsive external shell. These three configurations are displayed in the upper panels of Figure~\ref{fig3}a, where coumarin is shown as red beads; in simulations, we focus on microgels with a $10\%$ coumarin content. The corresponding radial density profiles for the different coumarin distributions are reported in Figure~\ref{fig3}b. At this stage, we mimic the effect of experimental irradiation by allowing coumarin-type beads to form bonds whenever two of them are found within a prescribed distance, as described in the Methods. This procedure introduces additional crosslinkers within the microgel network, reproducing the structural consequences of coumarin dimerization. The resulting post-irradiation microgel configurations are shown in the bottom panels of Figure~\ref{fig3}. After equilibration, we can directly compute the corresponding form factors and density profiles, and compare them with the pre-irradiated structures, according to a protocol already established for other structural comparison between experimental data and microgel models~\cite{scotti2026structural}.

\begin{figure}[t!]
\centering
\includegraphics[scale=0.9]{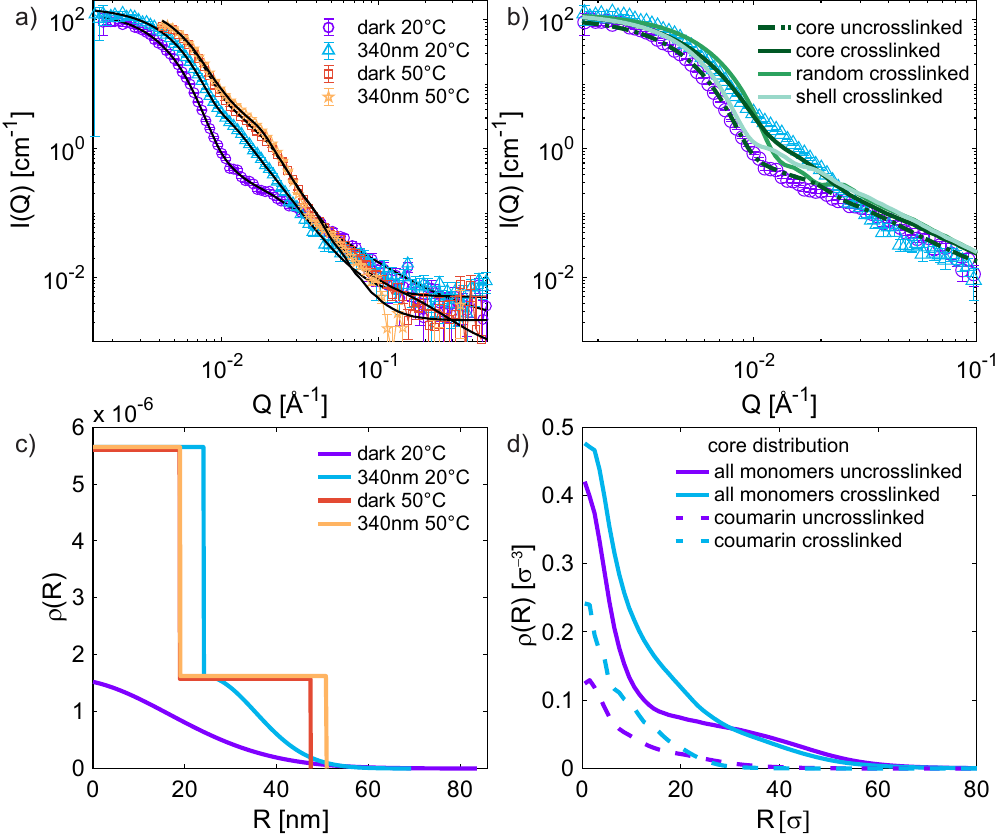}
\caption{\small 
Microgel form factors and density profiles as a function of light irradiation and temperature. a) SANS scattering intensities I(Q) of $\mu G_{10}$ in dark and after irradiation at 340 nm, at 20°C and 50°C, as indicated in the legend. Solid lines are fits according to eq. \ref{eq:core-fuzzy-shell}. b) Numerical form factors (solid lines) for microgels with crosslinked coumarin mainly distributed in the core, randomly or in the shell, as indicated in the legend, in comparison to the numerical form factor for a microgel with uncrosslinked coumarin (dash-dotted line) and to the experimental form factor (symbols) of $\mu G_{10}$ in dark and after irradiation at 340 nm at 20°C as in a). c) Polymer density distribution $\rho(R)$ as a function of the distance $R$ obtained from fitting of the experimental form factors in a). d) Numerical density profiles for (full lines) all monomers and (dashed lines) coumarin only of microgels with coumarin mainly distributed in the core, before and after crosslinking.}
\label{fig4}
\end{figure}

Experimentally, we performed SANS measurements on dilute microgel suspensions (0.1 w/w \%) in D$_2$O. In Figure \ref{fig4}a (purple circles) we report the form factor of $\mu G_{10}$ microgel in the dark at 20°C. After the Guinier region at low Q, the form factor displays a peak-less shape with a constant slope. This is consistent with star-like microgels obtained with a similar amount of EGDMA as crosslinker, \cite{Ballin2025} as well as after copolymerization with oligo (ethylene glycol) methyl ether methacrylate with an average $\text{M}_{\text{n}}$ = 950 g/mol. \cite{Bassu2025,Vialetto2026} 
We fit the form factor with the star-like core-shell model (eq.~\ref{eq:core-fuzzy-shell}), which depicts a particle comprising a highly crosslinked central core of radius $R_c$ and a fuzzy shell described by an extended star-like polymer density profile, modeled with the parameters thickness $t$ and fuzziness $\sigma$ (see more details in the Experimental section). Such a microgel conformation is attributable to the use of EGDMA due to its accumulation in the particle center during synthesis, resulting in a rigid core, a less crosslinked shell, and extended dangling chains with little to no crosslinking, which impart a star-like architecture to the particles.
Notably, this also suggests that such a peculiar conformation is a prerogative of EGDMA, as it is obtained with different monomers, from NIPAM to oligo ethylene glycols of varying chain length, and it is also maintained in the presence of coumarin comonomers.
Fit of $\mu G_{10}$ in the dark at 20°C with the star-like core-shell model yields a particle with $R_{tot}$ = 84 $\pm$ 4 nm, $R_c$ = 27.8 $\pm$ 0.6 nm and $2\sigma$ = 45 $\pm$ 1 nm (see also Figure S6). 

The form factor after extensive irradiation at 340 nm in Figure \ref{fig4}a (light blue triangles) shows a pronounced difference in the scattering profile at $Q \geq 10^{-2}$ \AA$^{-1}$, with the appearance of a shoulder at $Q \simeq 10^{-2}$ \AA$^{-1}$ and an increased slope at higher Q values. 
The changes we observe upon coumarin crosslinking are also consistent with previous results on star-like microgels that indicate the appearance of a shoulder to be linked to the formation of a more extended dense core and less densely pronounced star-like shell. \cite{Ballin2025}
Fitting with eq. \ref{eq:core-fuzzy-shell} reveals that the particle size decreases to $R_{tot}$ = 69 $\pm$ 11 nm. 
Importantly, there is a significant change in the internal structure with a collapse of the internal polymer network into a large dense core ($R_c$ = 42 $\pm$ 6 nm) surrounded by a thin shell of fuzziness $2\sigma$ = 22.6 $\pm$ 0.8 nm. The sharp increase of the exponent $\mu$ as well (Figure S6) indicates collapse of the polymer into an overall denser microgel.

We now turn to assess which coumarin distribution best represents the experimental data before and after crosslinking by means of computer simulations.
The excellent agreement, after appropriate rescaling, between the experimental form factor and the form factor calculated from simulations without additional coumarin-induced crosslinking confirms the star-like architecture of the microgel, as shown in Figure~\ref{fig4}b. We verified that, in these conditions, the use of different distributions of coumarin combined with the solvophobic potential does not affect the structure of the particle with respect to a standard star microgel (see Figure S15). As a consequence, all three models considered for the coumarin distribution capture the experimental data in the dark state.

In Figure~\ref{fig4}b, we report the calculated form factors after crosslinking for the three models discussed before in Figure~\ref{fig3} (solid lines). The form factor that best reproduces experiments is the one corresponding to a microgel with coumarin mainly placed in the core of the particle, from the center up to around half its extension. In the other cases, instead, the changes we observe at around $10^{-2}$ \AA$^{-1}$ are clearly not compatible with the experimental data. In particular, for a randomly distributed coumarin, although the overall size changes slightly after crosslinking, the internal structure does not appear to be significantly affected, as coumarin is too sparse within the particle. Similarly, when coumarin is localized in the outer region of the microgel, the shoulder appearing in the calculated form factor is not compatible with the experimental one. 
Beyond the reference case in which coumarin extends up to half of the particle radius, we also investigate slightly more confined and more extended distributions, reaching approximately $40\%$ and $55\%$ of the particle radius from the center, respectively, as reported in the SI (Figure S17).
Although all three models reproduce the experimental form factors reasonably well, a coumarin distribution that is too strongly concentrated in the microgel core (40\% of the particle radius) produces an excessively pronounced peak in the intermediate-to-high $Q$ range. In contrast, when coumarin is distributed more uniformly (55\% of the particle radius), the main effect is only a weaker shift of the form factor towards lower $Q$ values. Importantly, the coumarin beads must be considered to be solvophobic in order to correctly reproduce the experimental form factors (see Methods). In addition, the (marginal) effect of polydispersity is assessed in the SI (Figure S16).
Therefore, we conclude that the best agreement with experimental evidence is obtained when CMA is distributed throughout the inner half of the particle.
We note that this picture is also consistent with a previous work, where an indirect estimate based on microgel size growth and CMA incorporation during synthesis suggested that most CMA is located within the particle interior~\cite{Lu2019}.


The structural modifications within the microgels can be more clearly visualized by examining the radial density profiles in real space. In experiments, these profiles are obtained by extracting the parameters from fitting the form factors with eq. \ref{eq:core-fuzzy-shell}, whereas in simulations they can be computed directly from the simulation trajectory, as detailed in the Methods. The corresponding profiles are reported in Figure~\ref{fig4}c and~\ref{fig4}d, respectively, enabling a qualitative comparison between experiments and simulations.
In both cases, the additional crosslinking leads to an increase in polymer density in the central-to-inner region of the microgel, consistent with the picture emerging from the form factor analysis. 
For the simulations, we also report the density profiles of the coumarin beads alone (dashed lines in Figure~\ref{fig4}d). These profiles show that the increase in the overall microgel density is directly associated with bond formation in the central region of the particle, approximately for $R/\sigma < 20$. Indeed, the coumarin and total-density profiles display the same overall trend across the microgel, confirming that the structural rearrangement is mainly driven by the spatial localization and crosslinking of the coumarin moieties.

Furthermore, we explore the thermoresponsive behavior of the microgels through SANS experiments. The form factor of $\mu G_{10}$ in the dark at 50°C is reported in Figure \ref{fig4}a (red squares) for a direct comparison with the one at 20°C. Fitting with eq. \ref{eq:core-fuzzy-shell} displays the expected collapse above the microgel's VPTT, with a decrease of $R_{tot}$ and disappearance of the fuzziness (Figure S6 and \ref{fig4}c, red line). 
The scattering profile of $\mu G_{10}$ at 50°C after irradiation at 340 nm (orange stars) is superimposed to that in dark at the same temperature, indicating that irradiated microgels maintain thermal responsiveness, with a decrease of $R_{tot}$ and $R_c$, and a disappearance of the fuzziness. Additionally, the collapsed particles have essentially the same internal structure both in the pristine state or after irradiation at 340 nm.

A detailed comparison with DLS data is hindered by the distinct contributions to light and neutron scattering by the different regions of the polymer network. In the case of a star-polymer profile, in the swollen state the low density of the sparse dangling chains on the particle periphery does not contribute significantly to neutron scattering, causing an underestimation of $R_{tot}$, while it influences the diffusion of the microgels and the resulting $R_h$, as already reported for similar microgels, \cite{Vialetto2026} or when comparing dynamic and static light scattering. \cite{Bergman2020} As a result, the significant size decrease after irradiation of $\mu G_{10}$ is not captured in the form factor fits. However, the shell collapse and increase in $R_c$ and $\mu$ is consistent with a general collapse of the polymer network after light-induced crosslinking. Above the VPTT, the difference with DLS could instead be ascribed to heterogeneities in the outer microgel corona, stemming from the light-induced coumarin dimerization performed at high temperature when the dangling chains are collapsed onto the core. After re-swelling, such heterogeneities hamper complete microgel collapse due to the decreased ability of the entangled polymer network to rearrange (Figure \ref{fig2}c); however, being on the external fuzzy surface, they are not visible in SANS.


SANS of $\mu G_{5}$ (Figure S7) shows that the particle size remains substantially unchanged at 20°C after irradiation at 340 nm. However, variations in the I(Q) profiles at $Q \geq 10^{-2}$ are comparable to that observed for $\mu G_{10}$ and are consistent with a transition from a particle with a small core and an extended shell of low density in dark at 20°C, to one characterized by a much larger core and limited fuzziness upon extensive irradiation at 340 nm. This indicates that, while the particles size variation is of a lower extent due to the decreased amount of CMA, the internal polymer density profile can be tuned upon light irradiation, with a transition from a more diffuse to a denser network.
A sightly different light-responsive behavior is instead observed for $\mu G_{nip1}$ (Figure S8), for which irradiation at 340 nm induces a smaller decrease of the fuzziness, with the particle maintaining a pronounced star-like morphology even after extensive irradiation. The behavior at 50°C is instead comparable to $\mu G_{5}$ and $\mu G_{10}$, with a significant polymer collapse, as captured by an increase in $R_c$ and pronounced decrease of $2\sigma$.

\section{Conclusions}

In this work, we combined experiments and numerical simulations to understand in detail how the polymer network of light- and thermo-responsive microgels rearranges in response to optical and thermal stimulation. Light-responsiveness is provided by copolymerization with a coumarin derivative (CMA) that undergoes dimerization upon irradiation at 340 nm, forming additional crosslinking points in the polymer network. DLS showed that the particle size in the swollen state at low temperature is dependent on the amount of CMA, which controls the number of additional crosslinks after light exposure. The microgel’s thermal responsivity after irradiation decreases and, interestingly, for CMA $>$ 2.5 mol \% the size at high temperature is higher than that of the non-irradiated microgels. This was attributed to the formation of CMA crosslinking points that restrict network rearrangement, thereby limiting the polymer’s ability to incorporate or expel solvent, in stark contrast to the addition of higher amounts of crosslinker during particle synthesis. Similar results were obtained with different main monomers forming the polymer network, from the more conventional PNIPAM, to oligo (ethylene glycol) methyl ether methacrylates of different chain lengths, which are relevant for applications where a precise tuning of the microgel’s VPTT is of interest, as well as when biocompatibility is important.

SANS confirmed that these particles present a star-like polymer density profile, with a dense core and sparse polymer arms, a peculiarity attributed to the use of EGDMA as non-light-responsive crosslinker, \cite{Ballin2025,Vialetto2026} now demonstrated also with other thermoresponsive monomers. After light irradiation we observed the collapse of the external polymer chains into an overall denser microgel, which maintains a core-shell architecture with a larger core size and decreased corona, as observed in the density profiles. The modest size decrease after irradiation in SANS was ascribed to the limited contribution to neutron scattering, before irradiation, of the sparse, low-density dangling chains in the particle periphery. \cite{Vialetto2026} 
Numerical simulations corroborated the star-like architecture before irradiation. A comparison between experimental form factors and numerical ones obtained introducing hydrophobic CMA beads in different parts of the network revealed that, by using a batch synthesis process, the CMA comonomer accumulates mainly in the core of the microgels up to approximately half of the total particle radius. We anticipate that further experiments exploiting SANS with contrast variation in the presence of deuterated comonomers could support this finding. This information is especially relevant for applications where light can induce interparticle crosslinking in dense states where microgels might be interpenetrated.
Overall, a detailed understanding of the network conformation in thermal and light-responsive microgels is of paramount importance for translating their peculiar properties and performances into functional materials for applications spanning drug delivery, additive manufacturing, interfacial stabilization and responsive membranes. \cite{Das2024}


\begin{acknowledgement}

The authors acknowledge ISIS Neutron and Muon Source for beamtime allocation at the SANS2D and Larmor beamlines under proposals RB2410165 (https://doi.org/10.5286/ISIS.E.RB2410165-1) and RB2420121 (https://doi.org/10.5286/ISIS.E.RB2420121-1).
The authors gratefully acknowledge the financial support of Consiglio Nazionale delle Ricerche within CNR-STFC Agreement 2021-2027 (N 0065606), concerning collaboration in scientific research at the ISIS Neutron and Muon Source (UK) of Science and Technology Facilities Council (STFC).
J.V. acknowledges funding from Ministero dell'Università e della Ricerca (D.D. 247 published on 19.08.2022, grant Nr. MSCA\_0000004), funded by European Union - NextGenerationEU - PNRR, Missione 4, Componente 2, Investimento 1.2 and from Ministero dell'Università e della Ricerca through the Rita Levi Montalcini research program (D.M. 1317 published on 15-12-2021, grant Nr. PGR21W3GY8). M.L. and E.Z. acknowledge financial support by Progetto Co-MGELS funded by the European Union - NextGeneration EU under the National Recovery and Resilience Plan Mission 4 “Istruzione e Ricerca” - Component C2 - Investment 1.1 - “Fondo PRIN”, Project code PRIN2022PAYLXW Sector PE11, CUP B53D23008890006.

\end{acknowledgement}

\begin{suppinfo}

The Supporting Information files contains additional experimental characterizations, fitting parameters, and form factors of microgel models.

\end{suppinfo}

\bibliography{main}

\newpage

\begin{Large}
\section{Supporting information for:}
\end{Large}

\bigskip

\begin{center}
\begin{LARGE}
\textbf{Resolving Light-Induced Structural Rearrangements in Responsive Microgels}
\end{LARGE}
\end{center}

\setcounter{figure}{0}
\renewcommand{\thefigure}{S\arabic{figure}}
\renewcommand{\thetable}{S\arabic{table}}

\newpage

\begin{figure}[!htb]
\centering
\includegraphics[scale=0.6]{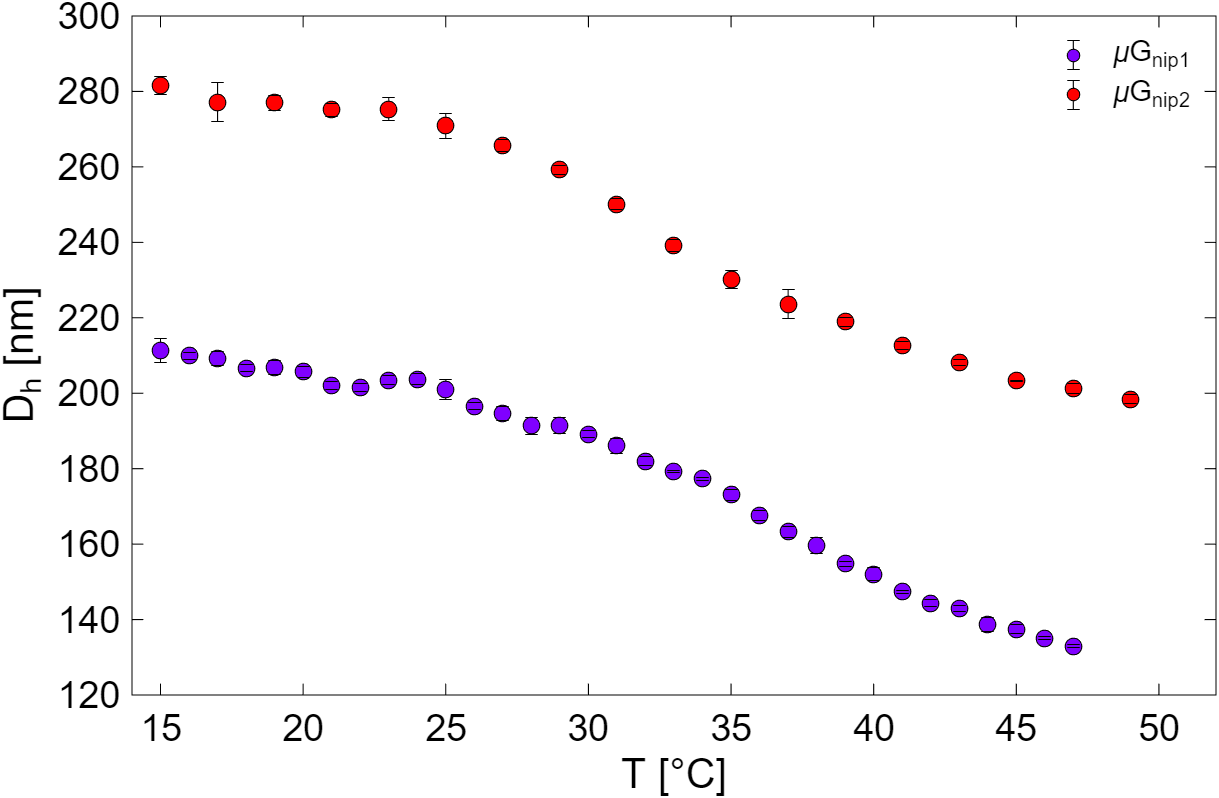}
\includegraphics[scale=0.6]{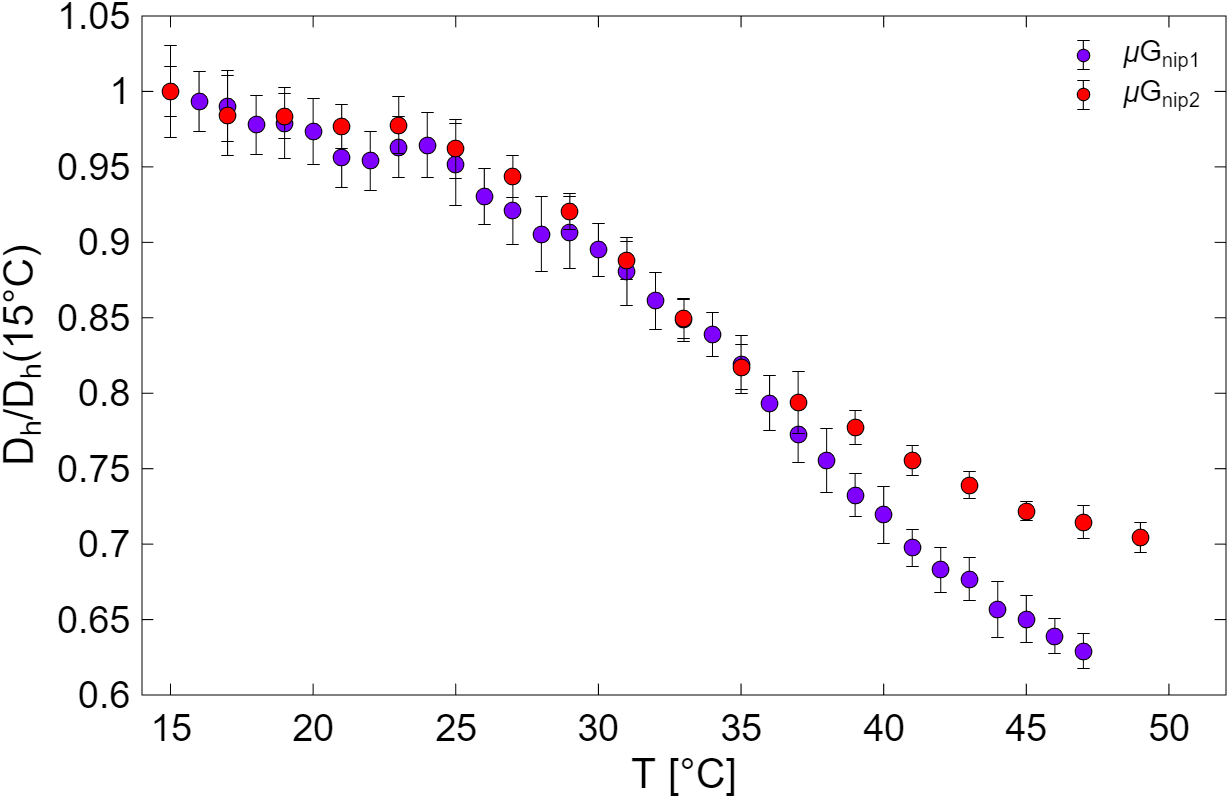}
\caption{\small \textbf{$D_h$ as a function of temperature for $\mu G_{nip1}$ and $\mu G_{nip2}$ in dark.} Bottom: diameter normalized with diameter at 15°C.}
\label{fig:nipam_DLS}
\end{figure}

\begin{figure}[!htb]
\centering
\includegraphics[scale=0.6]{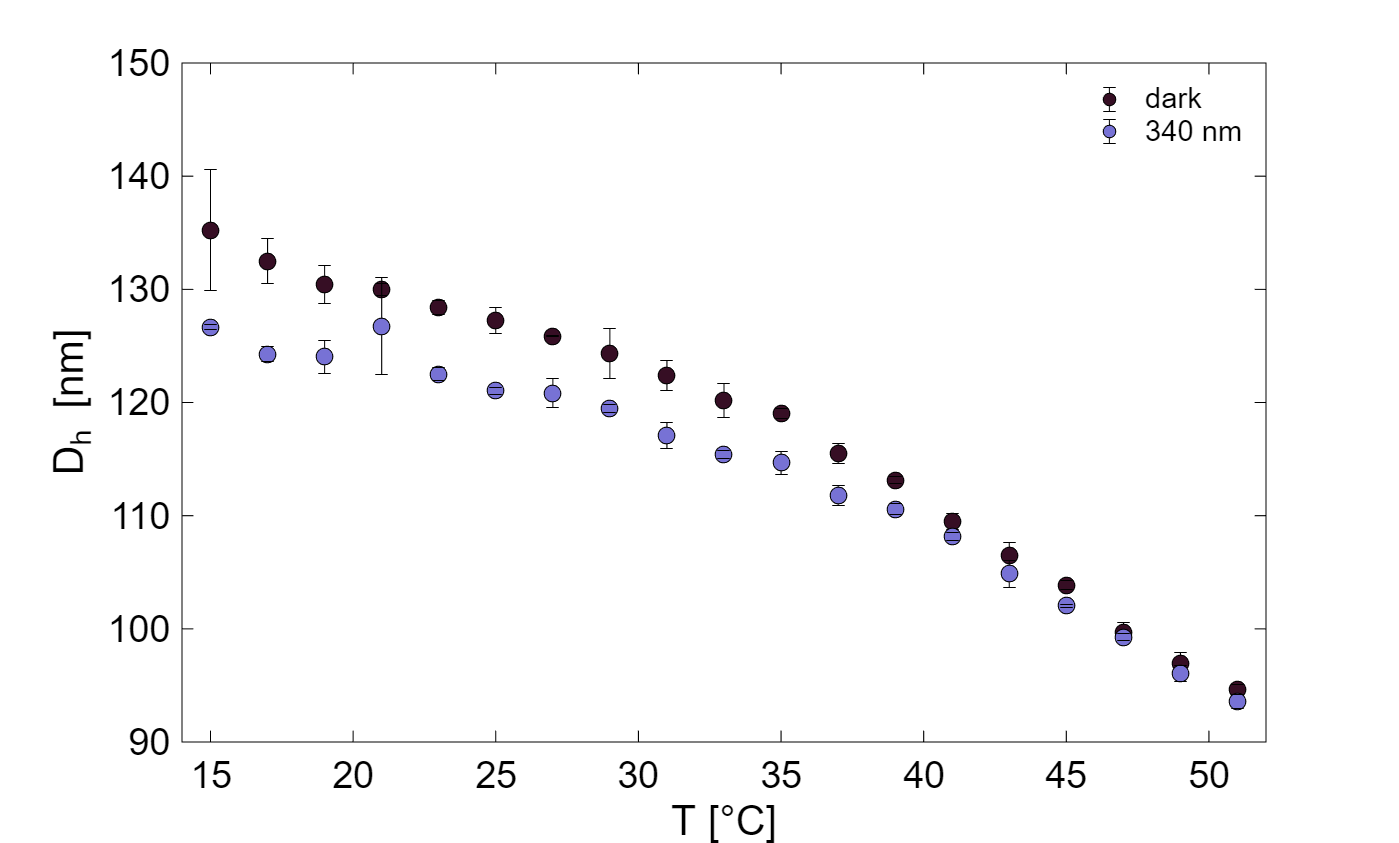}
\caption{\small \textbf{$D_h$ as a function of temperature for $\mu G_{2.5}$ in dark and after irradiation at 340 nm.}}
\label{fig:PEG-Cu-19 DLS}
\end{figure}

\begin{figure}[!htb]
\centering
\includegraphics[scale=0.6]{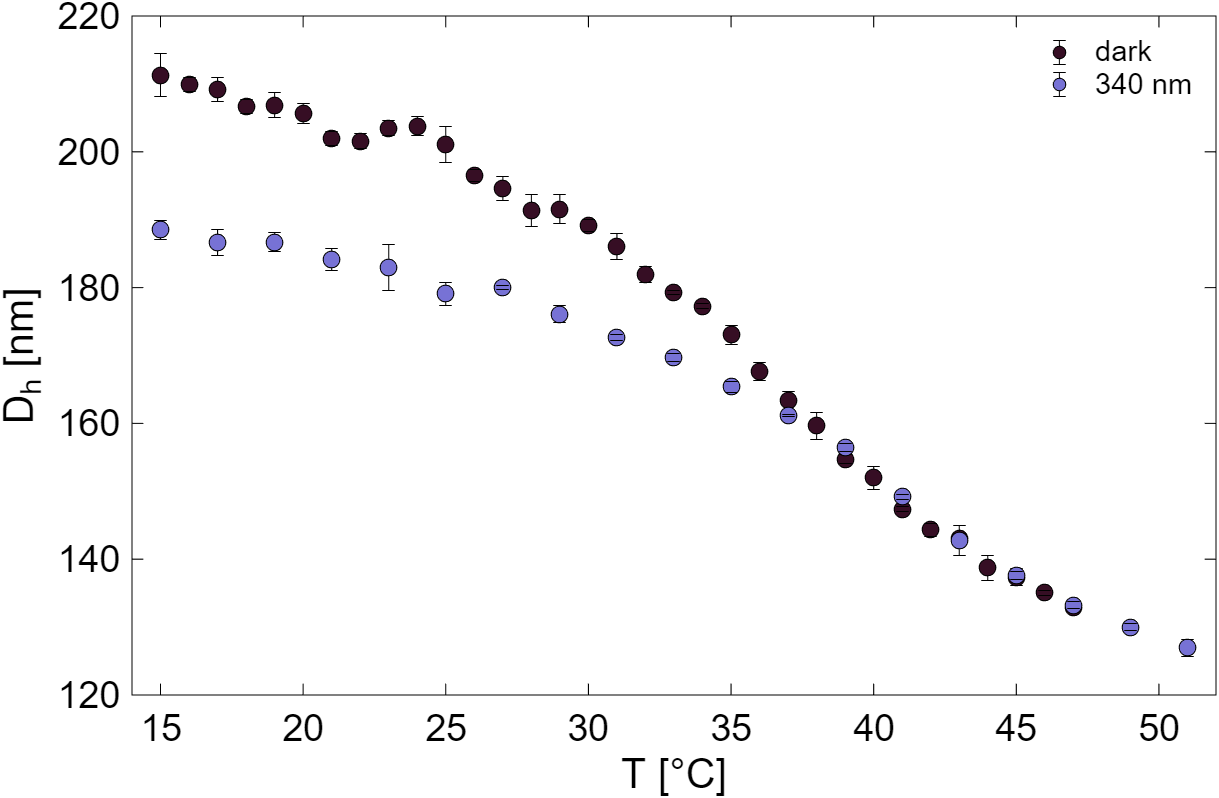}
\caption{\small \textbf{$D_h$ as a function of temperature for $\mu G_{nip1}$ in dark and after irradiation at 340 nm.}}
\label{fig:nipam_uGnip1 DLS}
\end{figure}

\begin{figure}[!htb]
\centering
\includegraphics[scale=0.6]{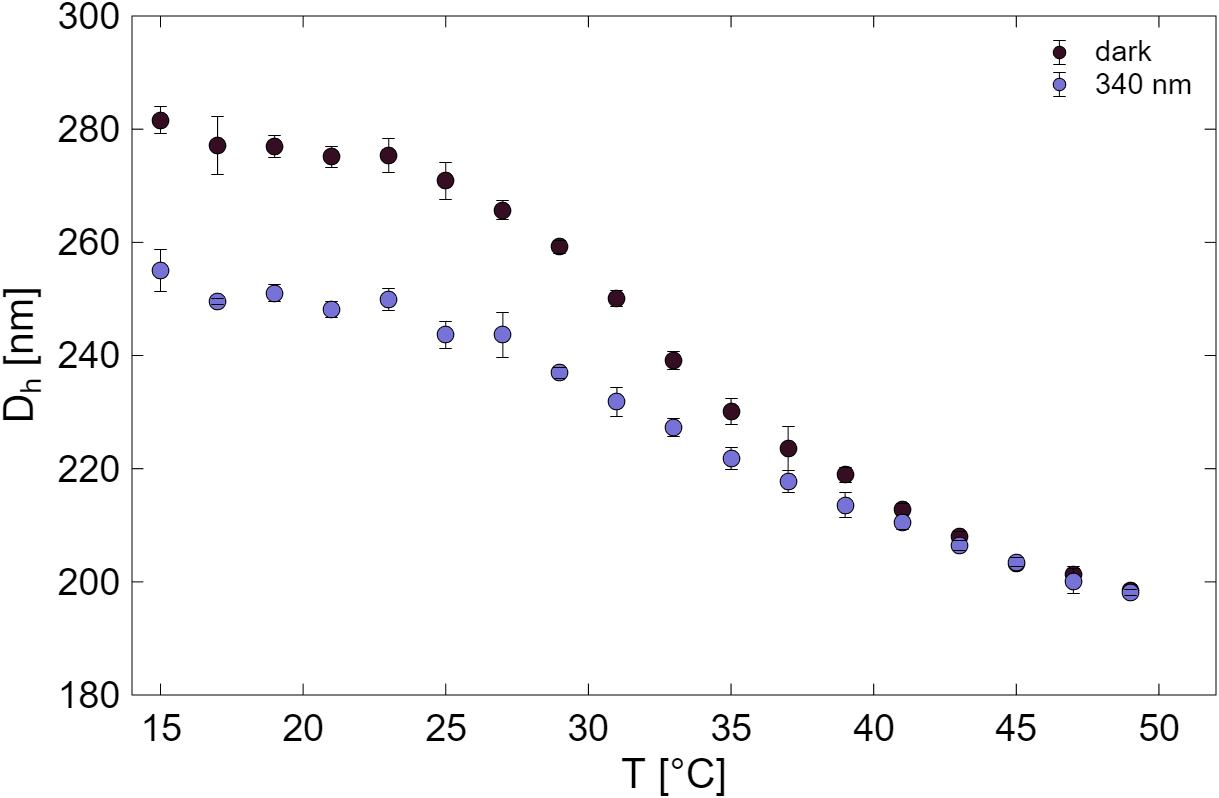}
\caption{\small \textbf{$D_h$ as a function of temperature for $\mu G_{nip2}$ in dark and after irradiation at 340 nm.}}
\label{fig:nipam_uGnip2 DLS}
\end{figure}

\begin{figure}[!htb]
\centering
\includegraphics[scale=0.6]{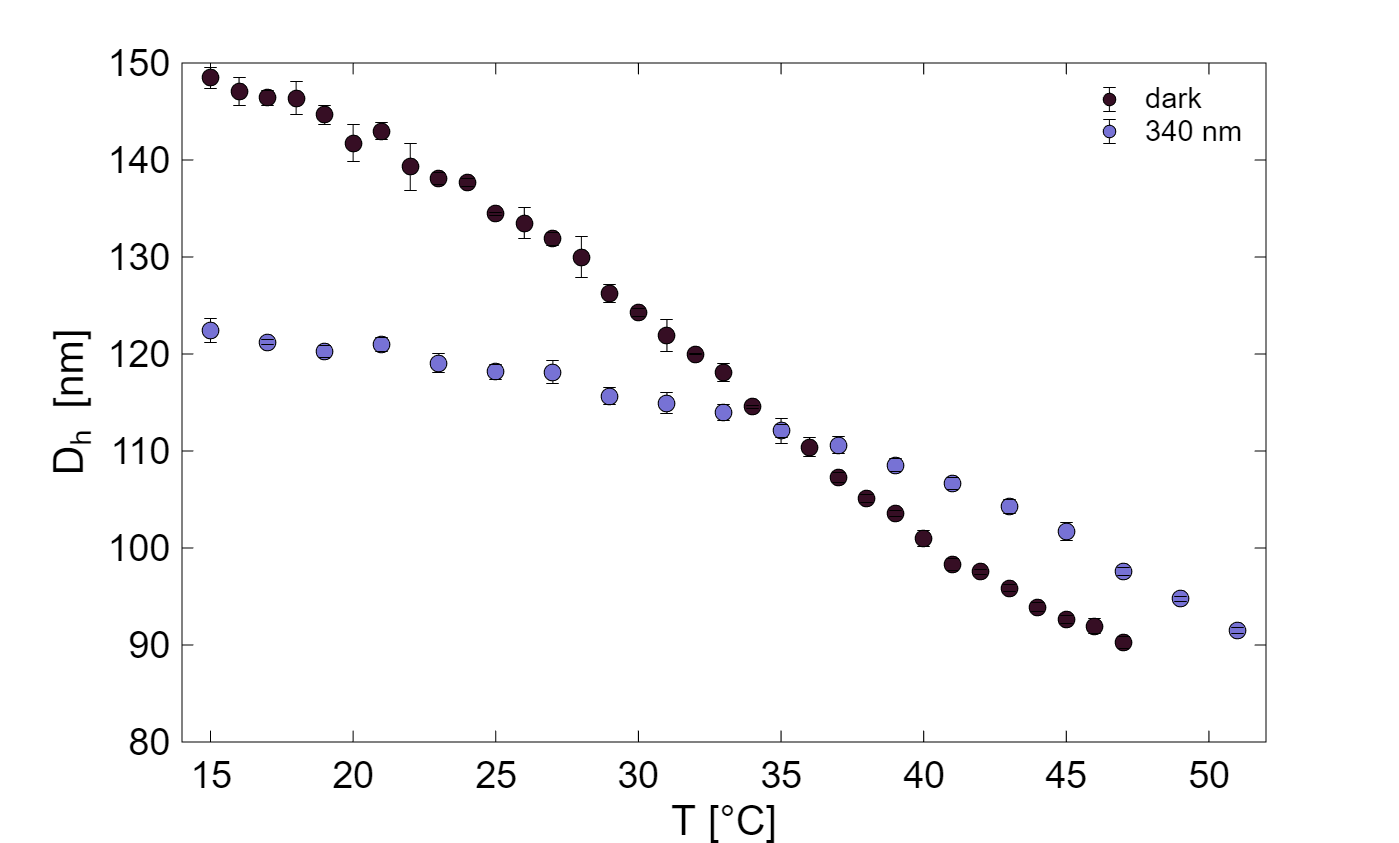}
\caption{\small \textbf{$D_h$ as a function of temperature for $\mu G_{5}$ in dark and after irradiation at 340 nm.}}
\label{fig:PEG-Cu-14 DLS}
\end{figure}

\begin{figure}[!htb]
\centering
\includegraphics[scale=1]{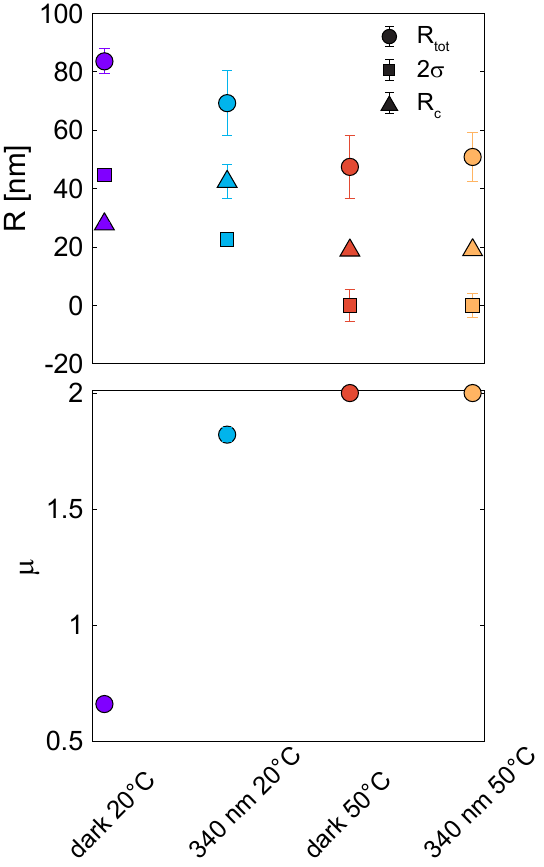}
\caption{\small \textbf{Parameters from star-like core-shell fits for $\mu G_{10}$.} Top: size parameters $R_{tot}$, $R_c$, and $2\sigma$. Bottom: parameter $\mu$.}
\label{fig:Figure 3 param}
\end{figure}

\begin{figure}[!htb]
\centering
\includegraphics[scale=0.92]{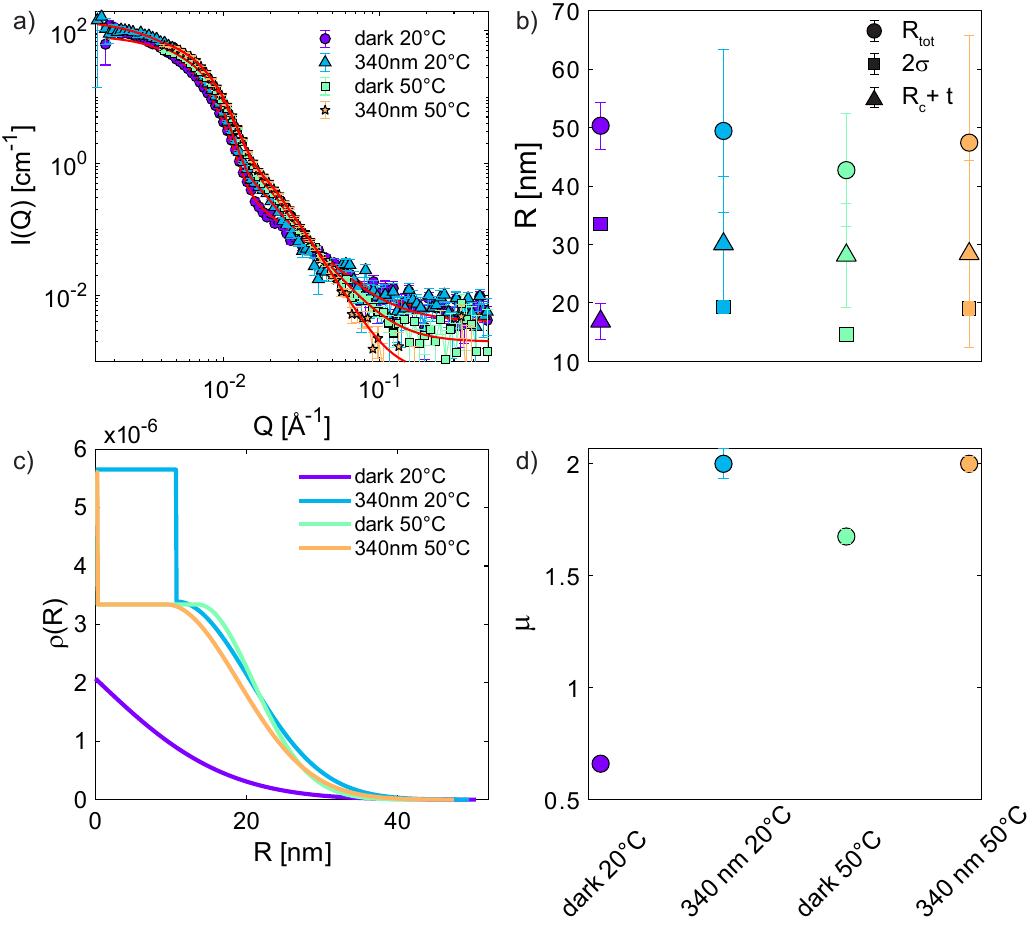}
\caption{\small \textbf{SANS of $\mu G_{5}$ as a function of light irradiation and temperature.} a) Scattering intensities I(Q) of $\mu G_{5}$ in D$_2$O in dark and after irradiation at 340 nm, at 20°C and 50°C, as indicated. Lines represent fits of the microgel form factor (more details in the main text). b) Polymer density distribution. c) Size parameters $R_{tot}$, $R_c + t$, and $2\sigma$. d) Parameter $\mu$.}
\label{fig:PEG-Cu-14 SANS}
\end{figure}

\begin{figure}[!htb]
\centering
\includegraphics[scale=0.92]{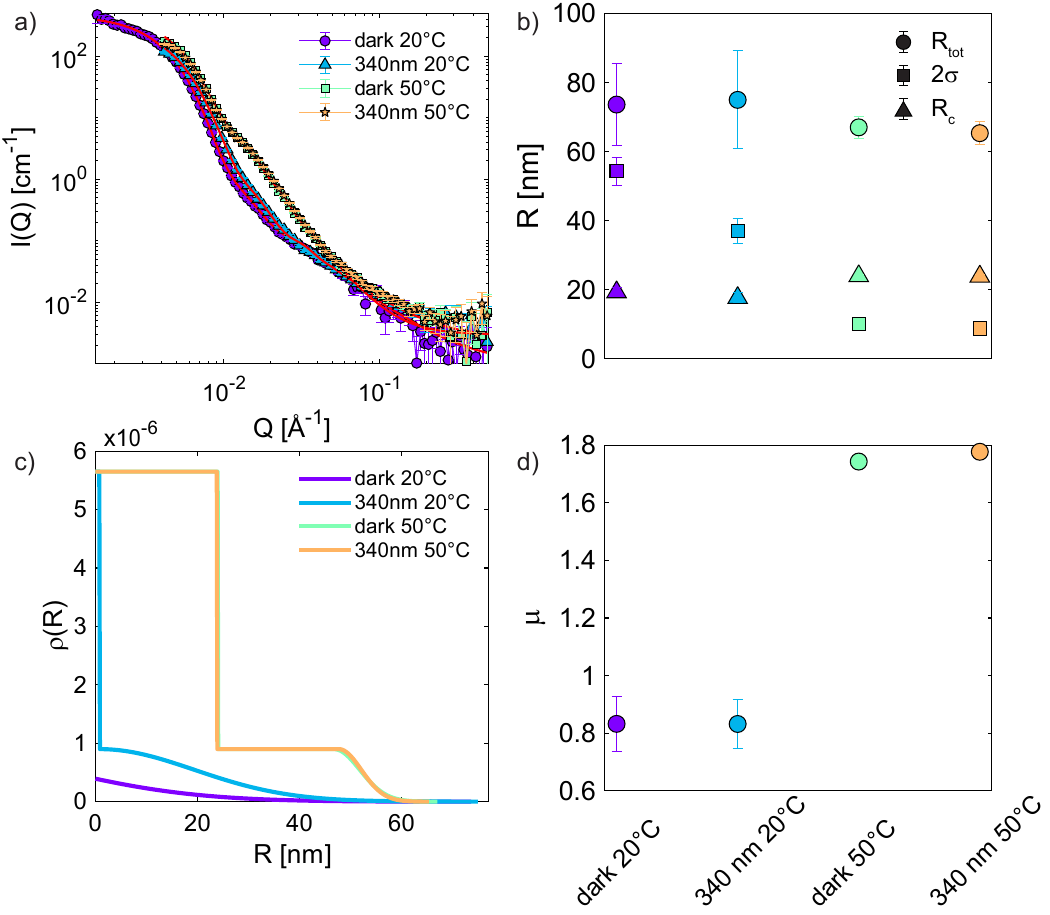}
\caption{\small \textbf{SANS of $\mu G_{nip1}$ as a function of light irradiation and temperature.} a) Scattering intensities I(Q) of $\mu G_{nip1}$ in D$_2$O in dark and after irradiation at 340 nm, at 20°C and 50°C, as indicated. Lines represent fits of the microgel form factor (more details in the main text). b) Polymer density distribution. c) Size parameters $R_{tot}$, $R_c$, and $2\sigma$. d) Parameter $\mu$.}
\label{fig:ME06 SANS}
\end{figure}

\begin{figure}[!htb]
\centering
\includegraphics[scale=1.8]{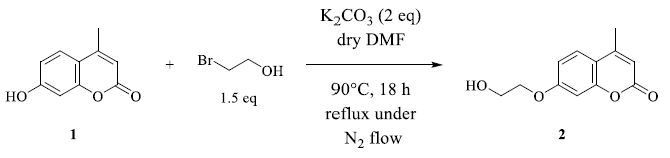}
\caption{\small \textbf{Synthesis of 7-(2-hydroxyethoxy)-4-methylcoumarin.}}
\label{fig:step1}
\end{figure}

\begin{figure}[!htb]
\centering
\includegraphics[scale=1.1]{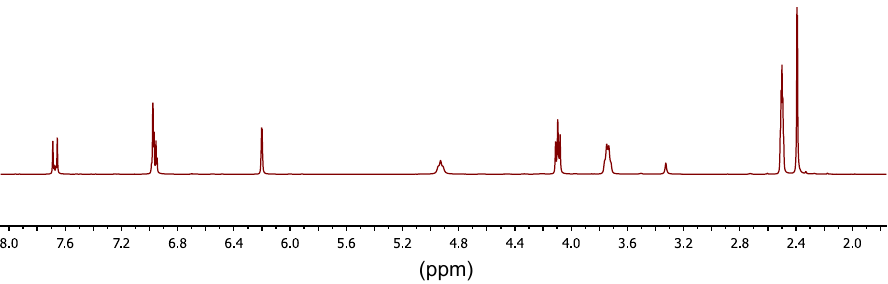}
\caption{\small \textbf{$^1H$ NMR of 7-(2-hydroxyethoxy)-4-methylcoumarin.} 1H NMR, 300 MHz in d6-DMSO.}
\label{fig:NMR1}
\end{figure}

\begin{figure}[!htb]
\centering
\includegraphics[scale=1.75]{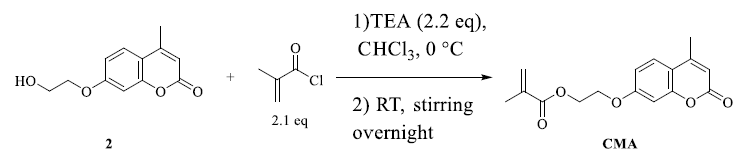}
\caption{\small \textbf{Synthesis of 2-((4-methyl-2-oxo-2H-chromen-7-yl)oxy)ethyl methacrylate (CMA).}}
\label{fig:step2}
\end{figure}

\begin{figure}[!htb]
\centering
\includegraphics[scale=1.1]{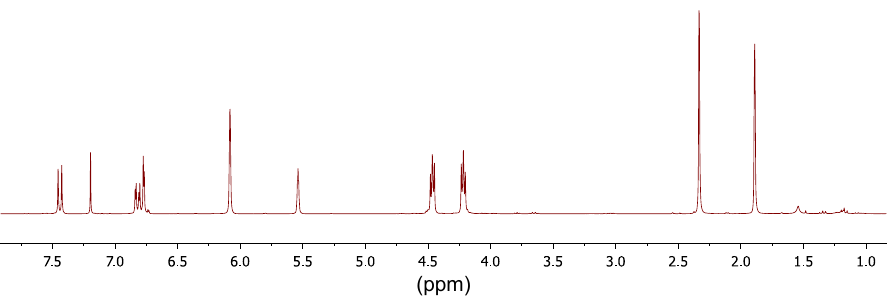}
\caption{\small \textbf{$^1H$ NMR of 2-((4-methyl-2-oxo-2H-chromen-7-yl)oxy)ethyl methacrylate (CMA).} 1H NMR, 300 MHz in CDCl$_3$.}
\label{fig:NMR2}
\end{figure}

\begin{figure}[!htb]
\centering
\includegraphics[scale=1.1]{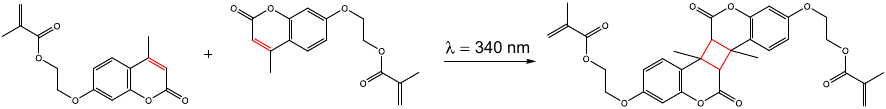}
\caption{\small \textbf{UV-induced CMA crosslinking.}} 
\label{fig:CMA_reaction}
\end{figure}

\begin{figure}[!htb]
\centering
\includegraphics[scale=0.9]{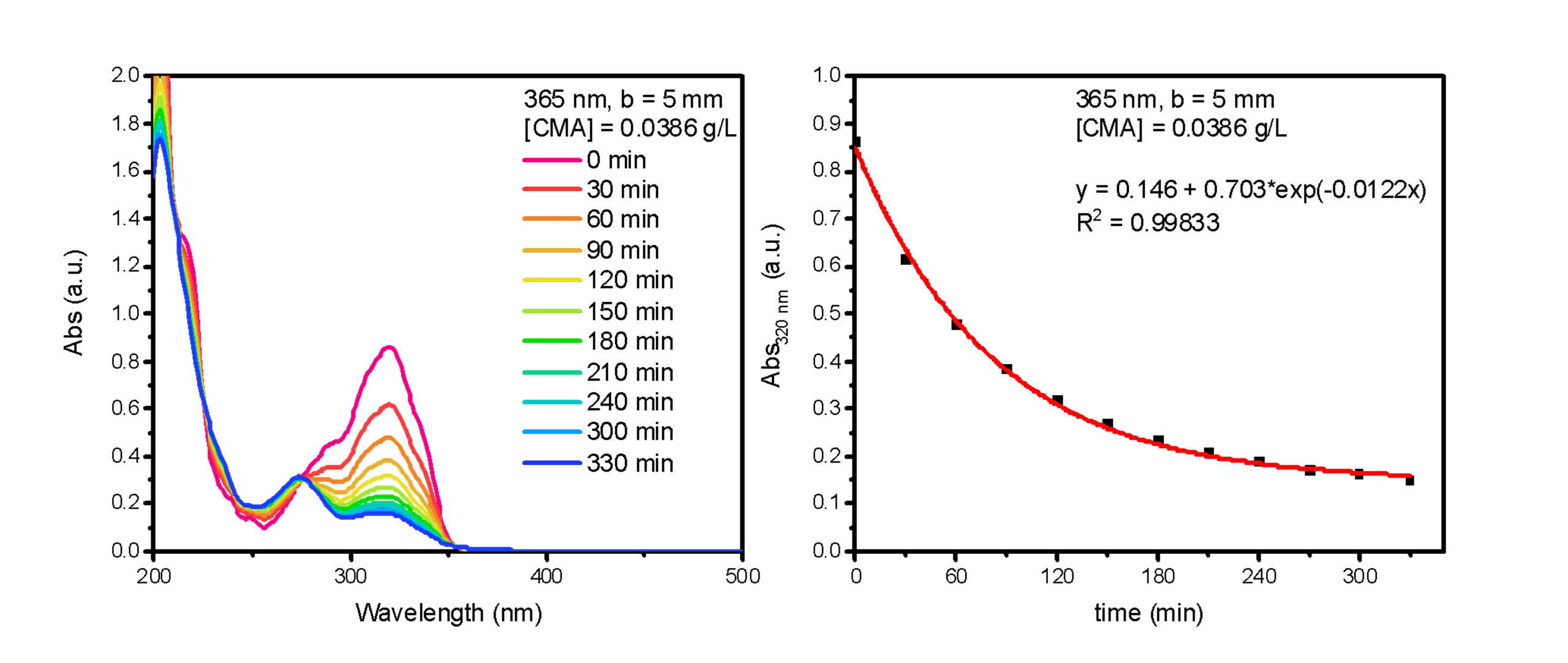}
\caption{\small \textbf{UV-vis spectroscopy of CMA.}}
\label{fig:CMA_irr}
\end{figure}

\begin{figure}[!htb]
\centering
\includegraphics[scale=0.5]{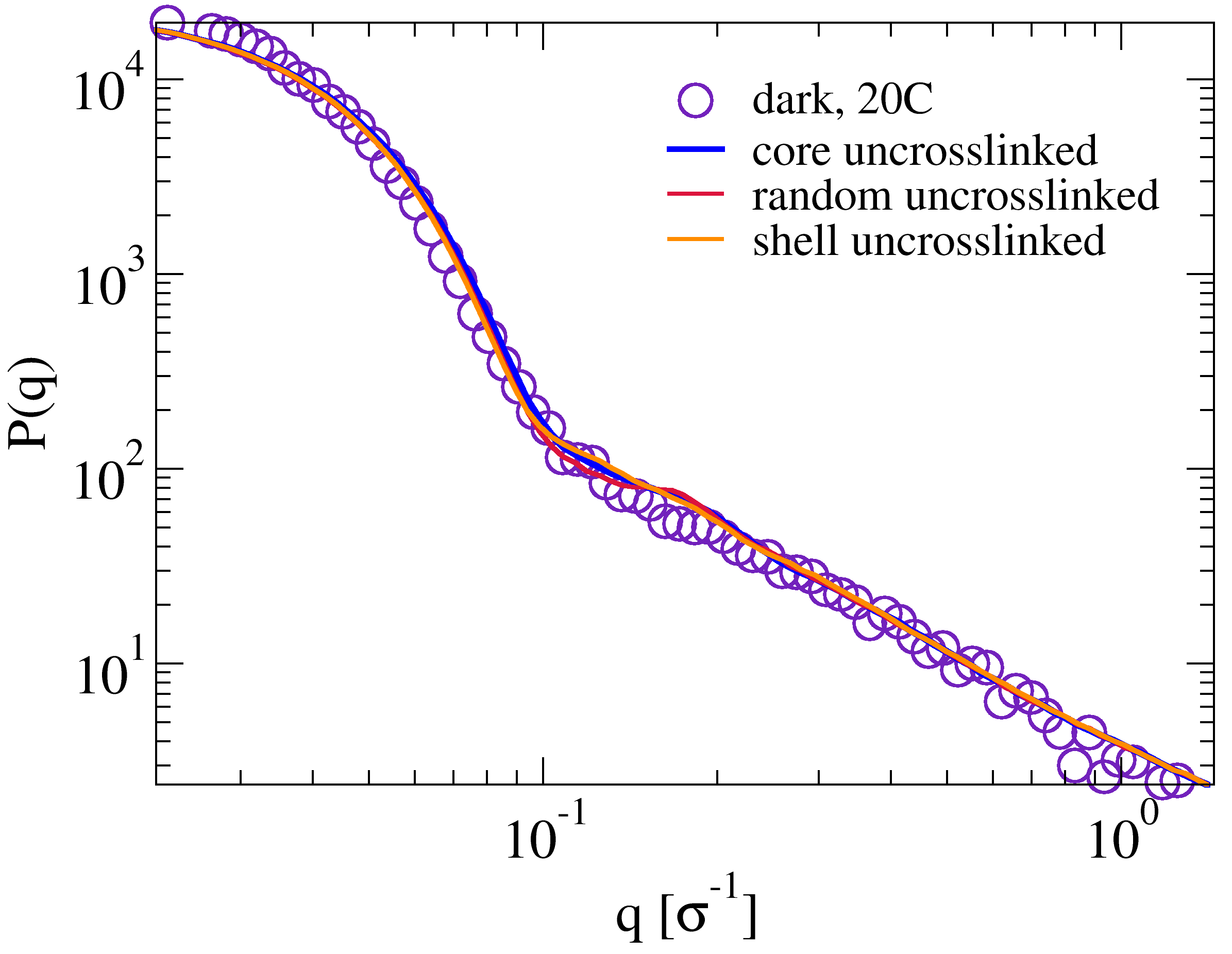}
\caption{\small \textbf{Comparison between different coumarin distributions.} Form factor $P(q)$ as a function of the wavevector $q$ for experiments in the dark and low temperature and simulations for three different coumarin distributions as described in the main text.}
\label{fig:cfr_sim_exp_dark}
\end{figure}

\begin{figure}[!htb]
\centering
\includegraphics[scale=0.7]{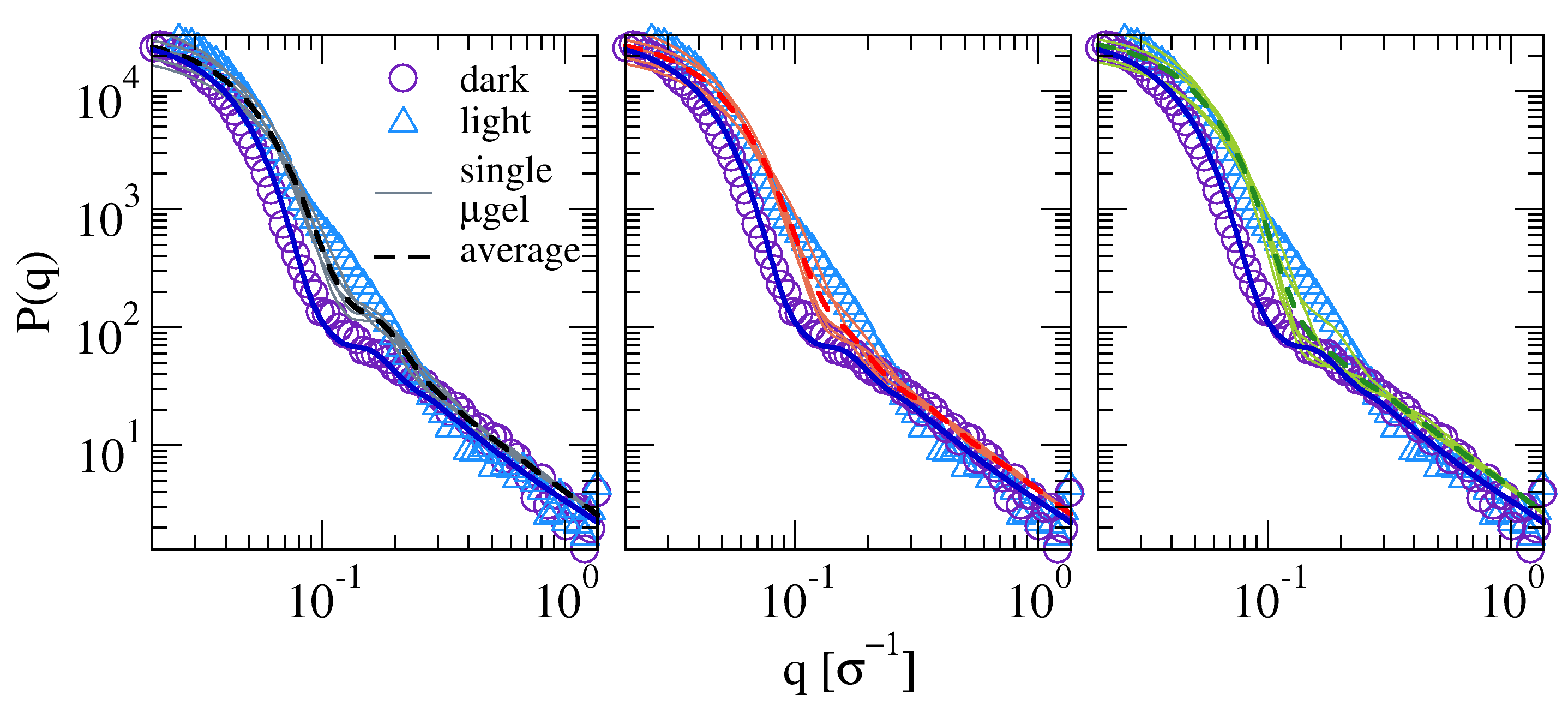}
\caption{\small \textbf{Effect of polydispesity on three different "core" coumarin distributions.} Form factor $P(q)$ as a function of the the wavevector $q$ for an increasing extension of the coumarin from the center of mass of the crosslinked microgel (from left to right) up to approximately $40, 50$ and $55\%$ of the whole microgel extension for microgels of different sizes (thin lines) and corresponding average (dashed thick line) as compared to the calculated form factor of the uncrosslinked microgel (full thick line) and experimental form factors (symbols). The corresponding coumarin density profiles are reported in Figure~\ref{fig:densprofcoum}.}
\label{fig:cfr_sim_exp_dark}
\end{figure}

\begin{figure}[!htb]
\centering
\includegraphics[scale=0.6]{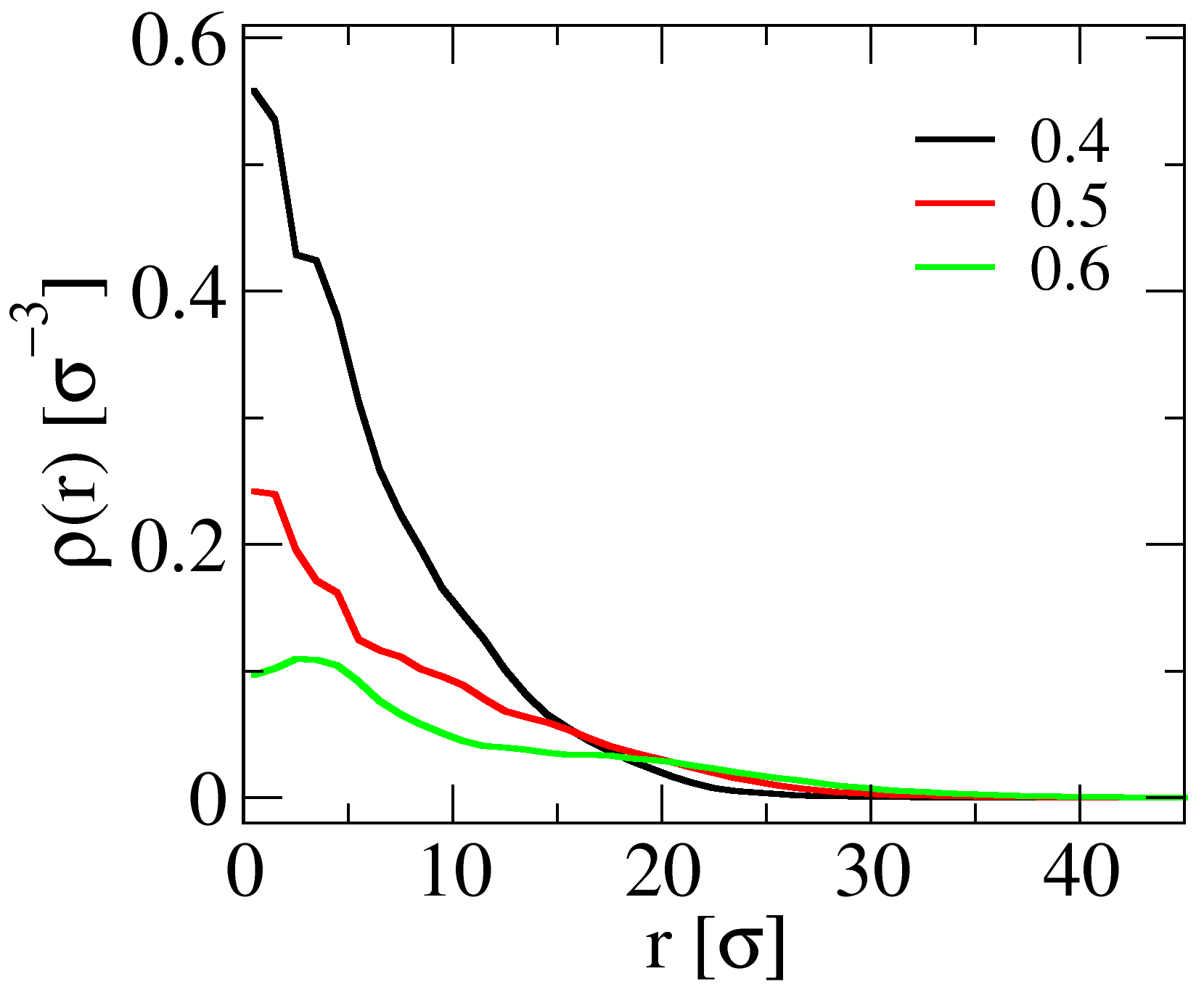}
\caption{\small \textbf{Coumarin density profiles.} Radial density profiles $\rho(r)$ as a function of the distance from the center of mass for microgels with coumarin distributed in the core for extension up to approximately $40, 50, 55\%$ of the whole microgel extension.}
\label{fig:densprofcoum}
\end{figure}

\end{document}